\documentclass[utf8]{frontiersinFPHY_FAMS} 

\setcitestyle{square} 
\usepackage{url,hyperref,lineno,microtype,subcaption}
\usepackage[onehalfspacing]{setspace}
\usepackage{bm}
\usepackage{braket}
\usepackage{amssymb}
\usepackage{graphicx}
\usepackage{dcolumn}
\usepackage{amsmath}
\usepackage{amsfonts}
\usepackage{xcolor}
\usepackage{hyperref}

\def\keyFont{\fontsize{8}{11}\helveticabold }
\def\firstAuthorLast{M.C. Atkinson} 
\def\Authors{M.C. Atkinson\,$^{1,*}$, W.H. Dickhoff\,$^{2}$}
\def\Address{$^{1}$Lawrence Livermore National Laboratory, Livermore, CA, USA\\
$^{2}$Department of Physics, Washington University in St. Louis, St. Louis, MO, USA}
\def\corrAuthor{M.C. Atkinson}

\def\corrEmail{mackenzie.c.atkinson@gmail.com}

\begin{document}
\onecolumn
\firstpage{1}

\title[Learning from knockout reactions using a dispersive optical model]{Learning from knockout reactions using a dispersive optical model} 

\author[\firstAuthorLast ]{\Authors} 
\address{} 
\correspondance{} 

\extraAuth{}

\maketitle

\
\begin{abstract}

   We present the empirical dispersive optical model (DOM) as applied to direct nuclear reactions. The DOM links both scattering and bound-state experimental data through a dispersion relation which allows for fully-consistent, data-informed predictions for nuclei where such data exists. In particular, we review investigations of the electron-induced proton knockout reaction from both $^{40}$Ca and $^{48}$Ca in a distorted-wave impulse approximation (DWIA) utilizing the DOM for a fully-consistent description. Viewing these reactions through the lens of the DOM allows us to connect the documented quenching of spectroscopic factors with increased high-momentum proton content in neutron-rich nuclei. 
   A similar DOM-DWIA description of the proton-induced knockout from $^{40}$Ca, however, does not currently fit in the consistent story of its electron-induced counterpart. With the main difference in the proton-induced case being the use of an effective proton-proton interaction, we suggest that a more sophisticated in-medium interaction would lead to consistent results.

\tiny
 \keyFont{ \section{Keywords:} nuclear structure, knockout reactions, optical potential, Green's function, distorted-wave impulse approximation} 
\end{abstract}

\section{Introduction}
\label{sec:intro}
Independent particle models (IPMs) provide a simplified picture of the nucleus in which correlations are neglected and all orbitals are 100\% filled up to the Fermi level according to the Pauli principle.
However, due to residual interactions there is depletion of orbitals below the Fermi energy and filling of those above it. Knockout reactions, in which a nucleon is removed from a nuclear target after collision with a projectile, are ideal tools for studying this distortion of the Fermi sea.
The importance of the $(e,e'p)$ reaction in clarifying the details of this rearrangement near the Fermi energy is well established and initially reviewed in Ref.~\cite{Advances14}. Subsequent high-resolution work at the Nikhef facility in Amsterdam then provided a detailed picture of the limitations of the IPMs in describing closed-shell nuclei~\cite{Gerard_12C,Denherder:1988,Steenhoven:1988,Kramer:1989,Peter90,Lex90,Ingo91,Lapikas93,Pandharipande97}.
The primary interaction in this reaction is electromagnetic and well understood so that at sufficiently high electron beam energy a distorted-wave impulse approximation (DWIA) provides an excellent reaction model~\cite{Boffi:1980,Giusti:1987,Giusti:1988,ElectroResponse}.

In the traditional application of the DWIA to $(e,e'p)$ cross sections, the Nikhef group typically utilized a global optical potential at the energy of the outgoing proton to describe the distorted wave.
The overlap function from the ground state to the relevant state in the nucleus with one proton removed was obtained from a standard Woods-Saxon potential with the depth adjusted to the separation energy and the radius constrained by the momentum dependence of the observed cross section.
The cross sections obtained for targets consisting of closed shell nuclei then typically require a scaling factor of 0.6 to 0.7 to generate an accurate description of the data~\cite{Lapikas93}.
This scaling factor, usually referred to as the (reduced) spectroscopic factor, corresponds to the normalization of the overlap function between the target
ground state and the excited state of the recoiling $A-1$ nucleus.
A spectroscopic factor less than one indicates a departure from the IPM. 
Furthermore, the data show that additional removal strength with essentially the same overlap function is located at nearby energies, providing clear evidence of
the fragmentation of the single-particle strength~\cite{Kramer:1989,Kramer90}. 

While stable targets corresponding to closed-shell nuclei have been investigated with the $(e,e'p)$ reaction, corresponding results for exotic nuclei are not available and may never be.
Alternative reactions have been explored in inverse kinematics at rare isotope facilities. For example, the heavy-ion nucleon knockout reaction was employed by researchers of the National Superconducting Cyclotron Laboratory at Michigan State University~\cite{Gade:2008,Gade:2014}.
The results suggested a strong dependence of the removal probability on the difference in separation energies of minority and majority species.
The analysis of these data for open-shell nuclei rely on small model space shell model calculations which already provide for partial occupation of orbits.
The resulting reduction factors for overlap functions similarly generated as for the $(e,e'p)$ reaction yield values close to 1 for the removal of valence majority nucleons and a strong suppression for the corresponding minority nucleons.
For closed-shell nuclei results were obtained with respect to the IPM description in line with the $(e,e'p)$ results mentioned above.
This dependence on nucleon asymmetry is not consistent with corresponding results of transfer reactions reviewed in Ref.~\cite{Dickhoff:2019} or the single-nucleon removal experiments recently reported in Refs.~\cite{Altar:2018,Kawase:2018}.
At this time no clear consensus has been reached on this intriguing difference.
A comprehensive status report of these different approaches containing a theoretical background was reported in Ref.~\cite{Aumann21}.

It has been argued in the literature that spectroscopic factors, while representing a useful concept, are not observables~\cite{Furnstahl02}.
No doubt the $(e,e'p)$ reaction provides the cleanest probe of removal probabilities. 
A similar approach in atoms for the $(e,2e)$ reaction backs up this claim (see \textit{e.g.} Ref.~\cite{Exposed!}).
Apart from assessing the accuracy of the DWIA method for the $(e,e'p)$ reaction it is also necessary to clarify the validity of the chosen ingredients of the Nikhef analysis.
We note that separate structure ingredients (phenomenological overlap function) and unrelated distorted waves obtained from local optical potentials were employed.
One approach to clarify these issues is provided by the dispersive optical model (DOM) first proposed by Mahaux and reviewed in Ref.~\cite{Mahaux91}.
In this article, we review the application of the DOM to DWIA calculations of knockout reactions~\cite{Atkinson:2018,Atkinson:2019,Yoshida:2021}.  
Recent implementations of the DOM have introduced fully nonlocal dispersive potentials~\cite{Mahzoon:2014,Dickhoff:2017} to allow additional data to be included in the description, like particle number and the nuclear charge density which were not considered in Ref.~\cite{Mahaux91}.
It is thus possible to extract all the ingredients of the DWIA for $(e,e'p)$ from the DOM description of all available elastic nucleon scattering data as well as separation energies, particle number, ground-state binding energy, charge radius, and the nuclear charge density for ${}^{40}$Ca and $^{48}$Ca in our case.
Indeed, the distorted outgoing proton wave and the overlap function with its implied normalization are all provided by the DOM to allow a consistent description of both $^{40}$Ca$(e,e'p)$$^{39}$K and $^{48}$Ca$(e,e'p)$$^{47}$K cross sections. 
The states analyzed for this reaction are the ground and first excited states of $^{39}$K and $^{47}$K, corresponding to the $0\textrm{d}3/2$ and $1\textrm{s}1/2$ valence hole states in the IPSM. 

The electron-induced proton knockout reaction, ($e$,$e'p$), has been considered the cleanest spectroscopic method for decades. 
An alternative approach is proton-induced knockout, or $(p,2p)$, which, despite some concerns about uncertainties~\cite{Jacob66,Jacob73,Chant77,Chant83,Samanta86,Cowley91,Wakasa17,Aumann21}, has been established as a complementary spectroscopic tool to ($e$,$e'p$) with about 15\% uncertainty for incident energy above 200~MeV~\cite{Wakasa17}. 
While the $(e,e'p)$ reaction involves one proton distorted wave, there are three such components in the $(p,2p)$ reaction.
In addition, the interaction responsible for the transition to the final state, apart from being fundamentally two-body in nature, involves an in-medium proton-proton ($pp$) interaction.
The $pp$ interaction is not nearly as well-understood as the electromagnetic transition operator ($ep$) in the $(e,e'p)$ reaction which is a predominantly one-body operator.

Using the same DOM ingredients that were employed in the DWIA analysis of $^{40}$Ca$(e,e'p)^{39}$K, we performed a DWIA calculation of $^{40}$Ca$(p,2p)^{39}$K in Ref.~\cite{Yoshida:2021}. This was not only the first implementation of consistent ingredients to this reaction, but also the first time that nonlocality in an optical potential was employed.
The resulting analysis pointed to an inconsistency between the electron-induced and proton-induced knockout reactions; while the DOM-provided spectroscopic factor of 0.71 perfectly describes the $(e,e'p)$ data, this factor had to be further reduced to 0.56 to reproduce the $(p,2p)$ data. Since the only difference between the descriptions of these two reactions is the $ep$ interaction vs. the $pp$ interaction, the inevitable conclusion is that a further study of the in-medium $pp$ interaction is required. 
We note that transfer reactions have also been studied with DOM ingredients~\cite{Nguyen11,Ross:2015,Potel17} but such studies require additional analysis of the reaction model although applying current non-local DOM potentials may provide useful insights.

The underlying Green's function ingredients of the single-particle propagator are presented in Sec.~\ref{sec:prop} while the DOM framework and its ingredients are introduced in Sec.~\ref{sec:DOM}.
The application to the $(e,e'p)$ reactions is described in Sec.~\ref{sec:DWIA}.
Results for the ${}^{40}$Ca$(e,e'p)$ and ${}^{48}$Ca$(e,e'p)$ reactions are presented in Secs.~\ref{sec:ca40-eep} and~\ref{sec:Ca48-eep}, respectively.
A discussion of the $(p,2p)$ results employing DOM ingredients is provided in Sec.~\ref{sec:p2p}.
Conclusions and some outlook are presented in Sec.~\ref{sec:conclusions}.

\section{Theory}
\label{sec:theory}
This section is organized to provide brief introductions into the underlying
theory of the DOM.

   \subsection{Single-particle propagator}
\label{sec:prop}
   The single-particle propagator describes the probability amplitude for adding (removing) a particle in state $\alpha$ at one time to (from) the non-degenerate ground state and propagating on top of that state until a later time when it is removed (added) in state $\beta$~\cite{Exposed!}.  In addition to the conserved orbital and
  total angular momentum ($\ell$ and $j$, respectively), the labels $\alpha$ and
  $\beta$ in Eq.~\eqref{eq:green} refer to a suitably chosen single-particle basis. 
  We employed a coordinate-space basis in our original $^{48}$Ca calculation in Ref.~\cite{Mahzoon:2017} but have since updated to using a Lagrange basis~\cite{Baye:2010} in all subsequent calculations (including that of $^{208}$Pb from Ref.~\cite{Atkinson20}).
 It is convenient to work with the Fourier-transformed propagator in the energy domain, 
 \begin{align}
    G_{\ell j}(\alpha,\beta;E)  &= \bra{\Psi_0^A}a_{\alpha \ell j}
    \frac{1}{E-(\hat{H}-E_0^A)+i\eta} a_{\beta \ell j}^\dagger\ket{\Psi_0^A}
    \nonumber \\ 
    + \bra{\Psi_0^A}&a_{\beta \ell j}^\dagger\frac{1}{E-(E_0^A-\hat{H})-i\eta}
    a_{\alpha \ell j}\ket{\Psi_0^A},
    \label{eq:green}
 \end{align}
 with $E^A_0$ representing the energy of the non-degenerate ground state $\ket{\Psi^A_0}$.
 Many interactions can occur between the addition and removal of the particle (or \textit{vice versa}), all of which need to be considered to calculate the propagator. 
 No assumptions about the detailed form of the Hamiltonian $\hat{H}$ need be made for the present discussion, but it will be assumed that a meaningful Hamiltonian exists that contains two-body and three-body contributions.
 Application of perturbation theory then leads to the Dyson equation~\cite{Exposed!} given by
 \begin{equation}
    G_{\ell j}(\alpha,\beta;E) = G_{\ell}^{(0)}(\alpha,\beta;E)  +
    \sum_{\gamma,\delta}G_{\ell}^{(0)}(\alpha,\gamma;E)\Sigma_{\ell
    j}^*(\gamma,\delta;E)G_{\ell j}(\delta,\beta;E) ,
    \label{eq:dyson}
 \end{equation}
 where $G^{(0)}_{\ell}(\alpha,\beta;E)$ corresponds to the unperturbed propagator (the propagator derived from the unperturbed Hamiltonian, $H_0$, which in the DOM corresponds to the kinetic energy)
 and $\Sigma_{\ell j}^*(\gamma,\delta;E)$ is the irreducible self-energy~\cite{Exposed!}. 
 The hole spectral density for energies below $\varepsilon_F$ is obtained from 
 \begin{equation}
    S^h_{\ell j}(\alpha,\beta;E) = \frac{1}{\pi}\textrm{Im}\ G_{\ell j}(\alpha,\beta;E), 
    \label{eq:spec}
 \end{equation}
 where the $h$ superscript signifies it is the hole spectral amplitude. For brevity, we drop this superscript for the rest of this review.
 The diagonal element of Eq.~\eqref{eq:spec} is known as the (hole) spectral function identifying the probability density for the removal of a single-particle state with quantum numbers $\alpha \ell j$ at
 energy $E$. The single-particle density distribution can be calculated from the hole spectral function in the following way, 
 \begin{equation}
    \label{eq:density}
    \rho_{\ell j}^{(p,n)}(r) = \sum_{\ell j} (2j+1) \int_{-\infty}^{\varepsilon_F}dE\ S^{(p,n)}_{\ell j}(r,r;E),
 \end{equation}
 where the $(p,n)$ superscript refers to protons or neutrons, and $\varepsilon_F = \frac{1}{2}(E^{A+1}_0 - E^{A-1}_0)$ is the average Fermi energy which separates the particle and hole domains~\cite{Exposed!}.
 The number of protons and neutrons ($Z,N$) is calculated by integrating $\rho_{\ell j}^{(p,n)}(r)$ over all space. In addition to particle number, the total binding energy can be calculated from the hole spectral function using the Migdal-Galitski sum rule~\cite{Exposed!}, 
 \begin{equation}
    E_0^{N,Z} = \frac{1}{2}\sum_{\alpha\beta}\int_0^{\varepsilon_F}dE\left[\braket{\alpha|\hat{T}|\beta}S^h(\alpha,\beta;E)
     + \delta_{\alpha\beta}ES^h(\alpha,\alpha;E)\vphantom{\braket{\alpha|\hat{T}|\beta}S^h(\alpha,\beta;E)}\right].
    \label{eq:energy_sumrule}
 \end{equation}
This expression assumes that the dominant contribution involves the two-nucleon interaction~\cite{Atkinson2020B,Atkinson2021}.

To visualize the spectral function of Eq.~\eqref{eq:spec}, it is useful to sum (or integrate) over the basis variables, $\alpha$, so that only an energy-dependence remains, $S_{\ell j}(E)$.
 The spectral strength $S_{\ell j}(E)$ is the contribution at energy $E$ to the occupation from all orbitals with angular momentum $\ell j$. It reveals that the strength for a shell can be fragmented, rather than isolated
 at the independent-particle model (IPM) energy levels.  
 \begin{figure*}[h]
    \begin{center}
       \includegraphics[width=0.49\linewidth]{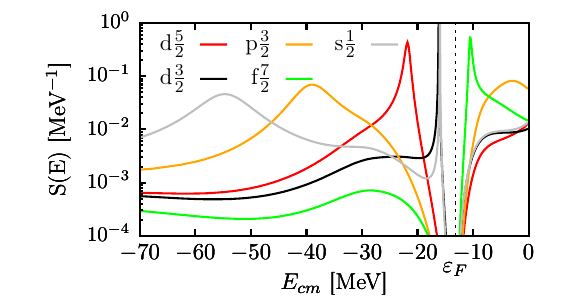}
       \includegraphics[width=0.49\linewidth]{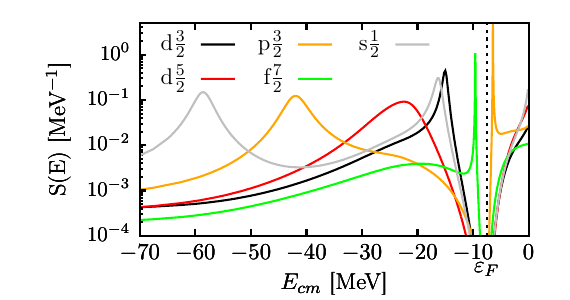}
    \end{center}
    \caption{\label{fig:spectral_n}Proton (left) and neutron (right) spectral functions of a representative set of $\ell j$ shells in $^{48}$Ca. The particle states are differentiated from the hole states by the dotted line representing $\varepsilon_F$. Figure adapted from Ref.~\cite{Atkinson:2019}.}

 \end{figure*}
 Figure~\ref{fig:spectral_n} shows the spectral strength for a representative set of proton (panel (a)) and neutron (panel (b)) orbits in $^{48}$Ca that would be considered bound in the IPM. 
 The  locations of the peaks in Fig.~\ref{fig:spectral_n} correspond to the energies of discrete bound states with one nucleon removed. For example, the s$1/2$ spectral function in Fig.~\ref{fig:spectral_n} has two peaks below
 $\varepsilon_F$ corresponding to the 0s$1/2$ and 1s$1/2$ quasihole states while the f$7/2$ spectral function has a peak below (neutrons) and above (protons) $\varepsilon_F$ corresponding to the 0f$7/2$ quasihole/particle state.  
 The wave functions of these quasihole/particle states can be obtained by transforming the Dyson equation into a nonlocal Schr\"{o}dinger-like equation by disregarding the imaginary part of $\Sigma^*(\alpha,\beta;E)$,
 \begin{align}
    \sum_\gamma\bra{\alpha}T_{\ell} + \textrm{Re}\ \Sigma^*_{\ell j}(\varepsilon_{\ell j}^n)\ket{\gamma}\psi_{\ell j}^n(\gamma) = \varepsilon_{\ell j}^n\psi_{\ell j}^n(\alpha),
    \label{eq:schrodinger}
 \end{align}
 where $\bra{\alpha}T_\ell\ket{\gamma}$ is the kinetic-energy matrix element, including the centrifugal term.
 The wave function, $\psi_{\ell j}^n(\alpha)$, is the overlap between the $A$ and $A-1$ systems and the corresponding energy, $\varepsilon_{\ell j}^n$, is the energy required to remove a nucleon with the particular quantum numbers $n\ell j$, 
 \begin{equation}
    \psi^n_{\ell j}(\alpha) = \bra{\Psi_n^{A-1}}a_{\alpha \ell j}\ket{\Psi_0^A}, \qquad \varepsilon_{\ell j}^n = E_0^A - E_n^{A-1}.
    \label{eq:wavefunction}
 \end{equation}

 When solutions to Eq.~\eqref{eq:schrodinger} are found near the Fermi energy where there is naturally no imaginary part of the self-energy, the normalization of the quasihole is well-defined as the spectroscopic factor, 
 \begin{equation}
    \mathcal{Z}^n_{\ell j} = \bigg(1 - \frac{\partial\Sigma_{\ell j}^*(\alpha_{qh},\alpha_{qh};E)}{\partial E}\bigg|_{\varepsilon_{\ell j}^n}\bigg)^{-1},
    \label{eq:sf}
 \end{equation}
 where $\alpha_{qh}$ corresponds to the quasihole state that solves Eq.~\eqref{eq:schrodinger}. The quasihole peaks in Fig.~\ref{fig:spectral_n} get narrower as the levels approach $\varepsilon_F$, which is a consequence of the imaginary part of the irreducible self-energy decreasing when
 approaching $\varepsilon_F$. In fact, the last mostly occupied neutron level in panel (b) of Fig.~\ref{fig:spectral_n} (0f$7/2$) has a spectral function that is essentially a delta function peaked at its energy level, where
 the imaginary part of the self-energy vanishes.  
 Valence proton hole orbits (1s$1/2$ and 0d$3/2$) exhibit the same behavior. 
 For these orbitals, the strength of the spectral function at the peak corresponds to the spectroscopic factor in Eq.~\eqref{eq:sf}. This spectroscopic factor is the very same we employ in the $(e,e'p)$ calculations discussed in Sec.~\ref{sec:ca40-eep} (see also Refs.~\cite{Atkinson:2018,Atkinson:2019}.

 \subsection{Dispersive optical model}
 \label{sec:DOM}

 The Dyson equation, Eq.~\eqref{eq:dyson}, simplifies the complicated task of calculating $G_{\ell j}(\alpha,\beta;E)$ from Eq.~\eqref{eq:green} to finding a suitable $\Sigma^*(\alpha,\beta;E)$ to invert.
 It was recognized long ago that $\Sigma^*(\alpha,\beta;E)$ represents the potential 
 that describes elastic-scattering observables~\cite{Bell59}. 
 The link with the potential at negative energy is then provided by the Green's function framework as was realized by Mahaux and Sartor who introduced the DOM as reviewed in Ref.~\cite{Mahaux91}. 
 The analytic structure of the nucleon self-energy allows one 
 to apply the dispersion relation, which relates the real part of the self-energy at a given energy to a dispersion integral of its imaginary part over all energies.
 The energy-independent correlated Hartree-Fock (HF) contribution~\cite{Exposed!} is removed by employing a subtracted dispersion relation with the Fermi energy used as the subtraction point~\cite{Mahaux91}.
 The subtracted form has the further advantage that the emphasis is placed on energies closer to the Fermi energy for which more experimental data are available.
 The real part of the self-energy at the Fermi energy is then still referred to as the HF term, and is sufficiently attractive to bind the relevant levels at about the correct energies.
 In practice, the imaginary part is assumed to extend to the Fermi energy on both sides while being very small in its  vicinity.
 The subtracted form of the dispersion relation employed in this work is given by
 \begin{align}
    \textrm{Re}\ \Sigma^*(\alpha,\beta;E) &= \textrm{Re}\
    \Sigma^*(\alpha,\beta;\varepsilon_F) \label{eq:dispersion} \\ -
    \mathcal{P}\int_{\varepsilon_F}^{\infty} \!\! \frac{dE'}{\pi}&\textrm{Im}\
    \Sigma^*(\alpha,\beta;E')\left[\frac{1}{E-E'}-\frac{1}{\varepsilon_F-E'}\right] \nonumber
    \\ + \mathcal{P} \! \int_{-\infty}^{\varepsilon_F} \!\!
    \frac{dE'}{\pi}&\textrm{Im}\
    \Sigma^*(\alpha,\beta;E')\left[\frac{1}{E-E'}-\frac{1}{\varepsilon_F-E'}\right],
    \nonumber      
 \end{align}
 where $\mathcal{P}$ is the principal value. 
 The static term, $\textrm{Re}\Sigma^*(\alpha,\beta;\varepsilon_F)$, is denoted by $\Sigma_{\text{HF}}$ from here on. 
 Equation~\eqref{eq:dispersion} constrains the real part of $\Sigma^*(\alpha,\beta;E)$ by empirical information of the HF and imaginary parts which are closely tied to experimental data. 
 Initially, standard functional forms for these terms were introduced by Mahaux and Sartor who also cast the DOM potential in a local form by a standard transformation which turns a nonlocal static HF potential into an energy-dependent local potential~\cite{Perey:1962}.
 Such an analysis was extended in Refs.~\cite{Charity06,Charity:2007} to a sequence of Ca isotopes and in Ref.~\cite{Mueller:2011} to semi-closed-shell nuclei heavier than Ca.
 The transformation to the exclusive use of local potentials precludes a proper calculation of nucleon particle number and expectation values of the one-body operators, like the charge density in the ground state (see Eq.~\eqref{eq:density}). 
 This obstacle was eliminated in Ref.~\cite{Dickhoff:2010}, but it was shown that the introduction of nonlocality in the imaginary part was still necessary in order to accurately account for particle number and the charge density~\cite{Mahzoon:2014}.
 Theoretical work provided further support for this introduction of a nonlocal representation of the imaginary part of the self-energy~\cite{Waldecker:2011,Dussan:2011}.
 A review detailing these developments was published in Ref.~\cite{Dickhoff:2017}.

 \subsubsection{Functional Form of the DOM Self-Energy}

 We employ a nonlocal representation of the self-energy following
 Ref.~\cite{Mahzoon:2014} where $\Sigma_{\text{HF}}(\bm{r},\bm{r'})$ and
 $\textrm{Im}\ \Sigma(\bm{r},\bm{r'};E)$ are parametrized and the energy-dependence of the real part, $\textrm{Re}\ \Sigma(\bm{r},\bm{r'};E)$, is generated from the dispersion relation in Eq.~\eqref{eq:dispersion}. The HF term consists of a volume term, spin-orbit term,  and a wine-bottle-shape generating term~\cite{Brida11},
\begin{align}
\Sigma_{HF}(\bm{r},\bm{r}') = V_\textrm{vol}(\bm{r},\bm{r}') + V_\textrm{so}(\bm{r},\bm{r}') + V_\textrm{wb}(\bm{r},\bm{r}') + \delta(\bm{r}-\bm{r}') V_C(r),
\end{align}
where the Coulomb potential, $V_C(r)$, is also included.
The radial part of our potentials takes the following form, 
\begin{equation}
   V_\textrm{vol}\left( \bm{r},\bm{r}' \right) =  V^\textrm{vol}
   \,f \left ( \tilde{r},r^\textrm{HF}_{(p,n)},a^\textrm{HF} \right ) 
   H \left( \bm{s};\beta^\textrm{HF} \right),
   \label{eq:HFvol} 
\end{equation}
where $V^\textrm{vol}$ is a parameter that determines the depth of the potential and $r^\textrm{HF}_{(p,n)}$, $a^{\textrm{HF}}$, and $\beta^\textrm{HF}$ are parameters that control the shape of the Woods-Saxon form factor $f$ and Perey-Buck-shaped~\cite{Perey:1962} nonlocality $H$, 
\begin{equation}
   f(r,r_{i},a_{i})=\left[1+\exp \left({\frac{r-r_{i}A^{1/3}}{a_{i}}%
   }\right)\right]^{-1} \hspace{1cm}
   H \left( \bm{s}; \beta \right) = \exp \left( - \bm{s}^2 / \beta^2 \right)/ (\pi^{3/2} \beta^3),
\end{equation}
and
\begin{equation}
   \tilde{r} =\frac{r+r'}{2} \hspace{3cm} \bm{s}=\bm{r}-\bm{r}'.
\label{eq:com_coords}
\end{equation}
Nonlocality is introduced in a similar way for $V_\textrm{wb}(\bm{r},\bm{r}')$ and $V_\textrm{so}(\bm{r},\bm{r}')$; their explicit forms can be found in Ref.~\cite{Atkinson:2020}.
The imaginary self-energy consists of volume, surface, and spin-orbit terms, 
\begin{align}
   \textrm{Im}\Sigma(\bm{r},\bm{r}';E) &= 
   -W^{vol}_{0\pm}(E) f\left(\tilde{r};r^{vol}_{\pm};a^{vol}_{\pm}\right)H \left( \bm{s}; \beta^{vol}\right) \nonumber \\
   &+ 4a^{sur}_\pm W^{sur}_{\pm}\left( E\right)H \left( \bm{s}; \beta^{sur}\right)
   \frac{d}{d \tilde{r} }f(\tilde{r},r^{sur}_{\pm},a^{sur}_{\pm})  + \textrm{Im}\Sigma_{so}(\bm{r},\bm{r}';E),
   \label{eq:imnl}
\end{align}
where $W^{vol}_{0\pm}(E)$ and $W^{sur}_{\pm}\left( E\right)$ are energy-dependent depths of the volume and surface potentials, respectively, and the $\pm$ subscript indicates there are different forms used above and below the Fermi energy (see Ref~\cite{Atkinson:2020} for exact forms). When considering asymmetric nuclei, such as $^{48}$Ca, additional terms proportional to the asymmetry, $\alpha_\mathrm{asy} = \frac{N-Z}{A}$, are added to $\Sigma_\textrm{HF}(\bm{r},\bm{r}')$ and $\textrm{Im}\Sigma(\bm{r},\bm{r}';E)$ for a Lane-like representation~\cite{Lane62}.
These asymmetric terms introduce additional parameters describing both their radial shape and energy-dependent depths~\cite{Atkinson:2020}. See Refs.~\cite{Atkinson:2020,Atkinson:2019} for the full list of parameters used in $^{48}$Ca.

 As mentioned previously, it was customary in the past to replace nonlocal potentials by local, energy-dependent potentials~\cite{Mahaux91,Perey:1962,Fiedeldey:1966,Exposed!}. The introduction of an energy dependence alters the dispersive
 correction from Eq.~\eqref{eq:dispersion} and distorts the normalization, leading to incorrect spectral functions and related quantities~\cite{Dickhoff:2010}. Thus, a nonlocal implementation permits the self-energy to accurately 
 reproduce important observables such as charge density, particle number, and ground-state binding energy. 

 To use the DOM self-energy for predictions, the parameters of the self-energy are constrained through weighted $\chi^2$ minimization (using the Powell method~\cite{Numerical}) by measurements of elastic differential cross sections ($\frac{d\sigma}{d\Omega}$), analyzing powers ($A_\theta$), reaction cross sections ($\sigma_\textrm{react}$), total cross sections ($\sigma_\textrm{tot}$), charge density ($\rho_{\textrm{ch}}$), energy levels ($\varepsilon_{n \ell j}$), particle number, and the root-mean-square charge radius ($R_{\text{ch}}$).
The angular-dependence of $\Sigma(\bm{r},\bm{r}';E)$ is represented in a partial-wave basis, and the radial component is represented in a Lagrange basis using Legendre and Laguerre polynomials for scattering and bound states, respectively.
 The bound states are found by diagonalizing the Hamiltonian in
 Eq.~\eqref{eq:schrodinger}, the propagator is found by inverting the Dyson
 equation, Eq.~\eqref{eq:dyson}, 
 while all scattering calculations are done in the framework of $R$-matrix theory~\cite{Baye:2010}. 

 The reproduction of all available experimental data (see Refs.~\cite{Mahzoon:2014,Atkinson:2018,Mahzoon:2017,Atkinson:2019} for comparisons to training data) indicates that we have realistic self-energies of $^{40}$Ca and $^{48}$Ca capable of describing both bound-state and scattering processes. 
 A parallel DOM analysis of these and other nuclei was conducted using Markov Chain Monte Carlo (MCMC) to optimize the potential parameters employing the same experimental data and a very similar functional form but with a reduced number of parameters. All observables from this MCMC fit fell within one standard deviation of those presented above~\cite{Pruitt:2020,Pruitt:2020C}.

\section{DWIA description of the $(e,e'p)$ cross section}
\label{sec:DWIA}
In the past, $(e,e'p)$ cross sections obtained at Nikhef in Amsterdam have been successfully described by utilizing the DWIA.
This description is expected to be particularly good when kinematics are used that emphasize the longitudinal coupling of the excitation operator, which is dominated by a one-body operator.
The Nikhef group was able to fulfill this condition by choosing kinematic conditions in which the removed proton carried momentum parallel or antiparallel to the momentum of the virtual photon.
Under these conditions, the transverse contribution involving the spin and possible two-body currents is suppressed. Therefore, the process can be interpreted as requiring an accurate description of the transition amplitude connecting the resulting excited state to the ground state by a known one-body operator.
This transition amplitude is contained in the polarization propagator which can be analyzed with a many-body description involving linear response~\cite{Exposed!}.
Such an analysis demonstrates that the polarization propagator contains two contributions.
The first term involves the propagation of a particle and a hole dressed by their interaction with the medium, but not each other.
The other term involves their interaction. The latter term will dominate at low energy when the proton that absorbs the photon participates in collective excitations like surface modes and giant resonances. 

When the proton receives on the order of 100 MeV it is expected that the resulting excited state can be well approximated by the dressed particle and dressed hole excitation~\cite{Brand90}.
In fact, when strong transitions are considered, like in the present work, two-step processes have only minor influence~\cite{Kramer:1989,Gerard_12C}. 
This interpretation forms the basis of the DWIA applied to exclusive $(e,e'p)$ cross sections obtained by the Nikhef group.
The ingredients of the DWIA therefore require a proton distorted wave describing the outgoing proton at the appropriate energy and an overlap function with its normalization for the removed proton.
The distorted wave was typically obtained from a standard (local) global optical potential like Ref.~\cite{Schwandt:1982} for ${}^{40}$Ca.
The overlap function was obtained by adjusting the radius of a local Woods-Saxon potential to the shape of the $(e,e'p)$ cross section while adjusting its depth to the separation energy of the hole.
Its normalization was obtained by adjusting the calculated DWIA cross section to the actual data~\cite{Lapikas93}.  
Standard nonlocality corrections were applied to both the outgoing and removed proton wave functions~\cite{Perey:63}, in practice making the bound-state wave function the solution of a nonlocal potential.
We observe that such corrections are $\ell$-independent and therefore different from the nonlocal DOM implementation.

   In order to describe the $(e,e'p)$ reaction, the incoming electron, the electron-proton interaction, the outgoing electron, and the outgoing proton must be addressed. 
   The cross section is calculated from the hadron tensor, $W^{\mu\nu}$,  which contains matrix elements of the nuclear charge-current density, $J^\mu$~\cite{ElectroResponse}.  
   Using the DWIA, 
   which assumes that the virtual photon exchanged by the electron couples to the 
   same proton that is detected~\cite{Giusti:1988,Boffi:1980},
   the nuclear current can be written as
   \begin{equation}
      J^\mu(\bm{q}) = \int d\bm{r}e^{i\bm{q}\cdot\bm{r}}\chi^{(-)*}_{E\ell j}(\bm{r})(\hat{J}^\mu_{\text{eff}})_{E\ell j}(\bm{r})\psi^n_{\ell j}(\bm{r})\sqrt{\mathcal{Z}^n_{\ell j}},
      \label{eq:current}
   \end{equation}
   where $\chi^{(-)*}_{E}(\bm{r})$ is the outgoing proton distorted wave~\cite{ElectroResponse}, 
   $\psi^n_{\ell j}$ is the overlap function, $\mathcal{Z}^n_{\ell j}$ its normalization, $\bm{q} = \bm{k_f} - \bm{k_i}$ is the electron three-momentum transfer, and 
   $\hat{J}^\mu_{\text{eff}}$ is the effective current operator~\cite{ElectroResponse}. 
   The incoming and outgoing electron waves
   are treated within the Effective Momentum Approximation, where the waves are represented by plane waves with effective momenta to account for distortion from the interaction with the target 
   nucleus~\cite{Giusti:1987}
   \begin{equation}
      k_{i(f)}^{\text{eff}} = k_{i(f)} + \int d\bm{r}V_c(\bm{r})\phi_{\ell j}^2(\bm{r}),
      \label{eq:effective}
   \end{equation}
   where $V_c(\bm{r})$ is the Coulomb potential of the target nucleus.
   This alters Eq.~(\ref{eq:current}) by replacing $\bm{q}$ with $\bm{q}_{\text{eff}}$.

   In the plane-wave impulse approximation (PWIA), in which the outgoing proton wave is approximated by a free scattering (plane) wave, the $(e,e'p)$ cross section can be factorized into an off-shell electron-proton cross section and the spectral function~\cite{ElectroResponse}, 
   \begin{equation}
      S(E_m,\bm{p}_m) = \frac{1}{k\sigma_{ep}}\frac{d^6\sigma}{dE_{e'}d\Omega_{e'}dE_pd\Omega_p}.
      \label{eq:momdist}
   \end{equation}
   The off-shell electron-proton cross section, $\sigma_{ep}$, is approximated from the on-shell one using the $\sigma_{\text{cc1}}$ model as proposed in~\cite{deForest:1983}.
   The factorization in Eq.~\eqref{eq:momdist} does not hold true for the DWIA, but $(e,e'p)$ cross sections, both experimental and theoretical, are typically divided by $\sigma_{\text{cc1}}$ when displayed.
  In principle, corrections due to two-step processes could be considered but they are estimated to make negligible contributions for the transitions considered in this study~\cite{Kramer:1989}.
   
  The calculations of the $(e,e'p)$ cross sections in Ref.~\cite{Atkinson:2018} were performed by employing DOM ingredients that were constrained by the experimental data discussed in Sec.~\ref{sec:DOM}. Appropriate distorted waves and overlap functions with their normalization were thus generated that allow for a DWIA description of the exclusive $(e,e'p)$ cross section for valence holes in ${}^{40}$Ca. Agreement with cross sections therefore supports the description of the reaction in a DWIA framework, but also confirms the overall consistency of the DOM approach including its interpretation of the normalization of the overlap functions as spectroscopic factors that can be confronted with data.

\subsection{$^{40}$Ca$(e,e'p)^{39}$K}
\label{sec:ca40-eep}

\begin{figure}[t]
   \begin{center}
         \includegraphics[scale=0.7]{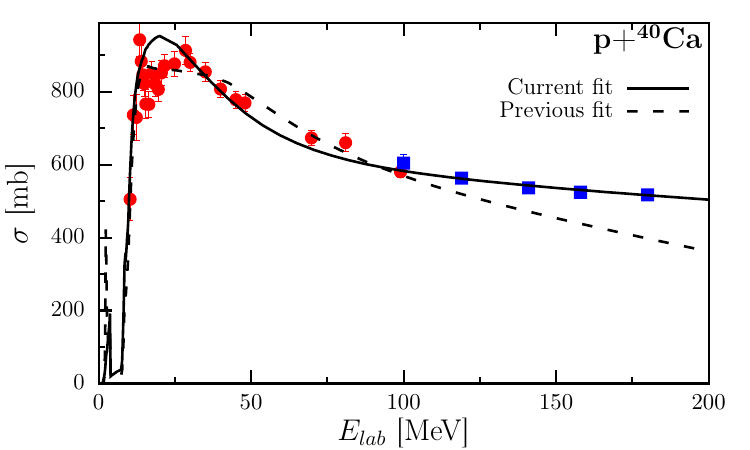}
      \end{center}
      \caption{The proton reaction cross section for $^{40}$Ca. The solid line represents the newest fit~\cite{Atkinson:2018}, while the dashed line depicts the original fit~\cite{Mahzoon:2014}. The circular data points were included in the original fit, while the square data points~\cite{PhysRevC.71.064606} were added in the newest fit. Figure adapted from Ref.~\cite{Atkinson:2018}}
   \label{fig:react}
\end{figure}

The first nonlocal DOM description of ${}^{40}$Ca data was presented in Ref.~\cite{Mahzoon:2014}.
In the mean time, additional experimental higher-energy proton reaction cross sections~\cite{PhysRevC.71.064606} had been incorporated which caused some adjustments of the DOM parameters compared to Ref.~\cite{Mahzoon:2014}. The updated parameters are collected in App. A of Ref.~\cite{Atkinson:2018}.
Adjusting the parameters from the previous values~\cite{Mahzoon:2014} to describe these additional experimental results leads to an equivalent description for all data except these reaction cross sections.
   These higher-energy data dictate that the proton reaction cross section stay flat for energies in the region around 150 MeV, as shown in Fig.~\ref{fig:react}. This means there is more absorption at higher energies than in the previous fit, leading to increased strength in the imaginary part of the self-energy. Due to the dispersion relation, Eq.~(\ref{eq:dispersion}), this increases the spectral strength at positive energies when the Dyson equation is solved. The sum rule discussed in detail in~\cite{Dussan:2014}, pertaining to the integral over all energies of the strength of the valence holes then implies that strength is transferred from below the Fermi energy to the energies with an increased imaginary part. 
   This resulting loss of strength below the Fermi energy reduces the spectroscopic factors by about 0.05 compared to the results reported in Ref.~\cite{Mahzoon:2014}.

\begin{figure}[b]
      \begin{center}
            \includegraphics{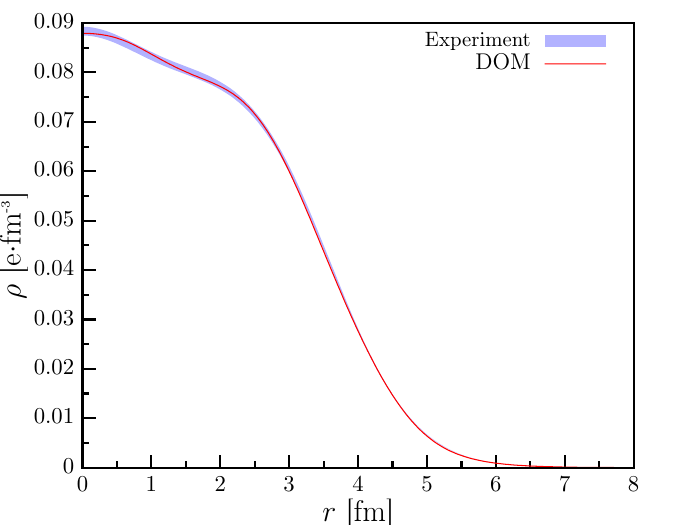}
         \end{center}
      \caption{Experimental and fitted $^{40}$Ca charge density. The solid line is calculated using the DOM propagator, 
      while the experimental band represents the 1\% error associated with the extracted charge density from elastic electron scattering experiments~\cite{deVries:1987,Sick79}. Figure adapted from Ref.~\cite{Atkinson:2018}.}
   \label{fig:chd}
\end{figure}

To accurately calculate the $(e,e'p)$ cross section in the DWIA, it is imperative that the DOM self-energy not only precisely generate available elastic scattering data but bound-state information as well. 
This is due to the fact that the shape of the cross section is primarily determined by the bound-state overlap function~\cite{Kramer:1989}. Thus, not only should the experimental charge radius be reproduced, but the charge density should match the experimental data, as we report in Fig.~\ref{fig:chd}, where the DOM charge density is shown as the solid line and compared with the deduced charge density (Fourier-Bessel parametrization) obtained from~\cite{deVries:1987} with the band representing the 1\% error.  

\begin{figure}[t]
   \begin{center}
         \includegraphics[width=\linewidth]{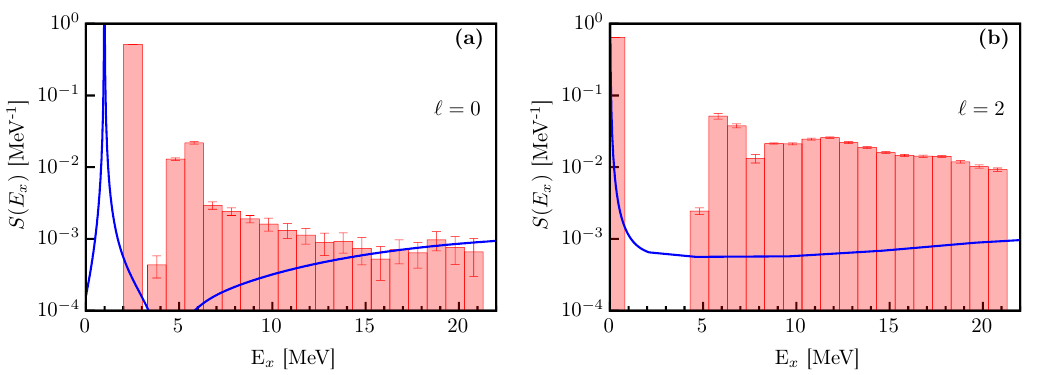}
      \end{center}
      \caption{ Spectral strength as a function of excitation energy for (a) the $1\textrm{s}1/2$ and (b) the $0\textrm{d}3/2$ proton orbitals, calculated from the DOM using Eq.~\eqref{eq:spec}) (solid line) 
   and extracted from the $^{40}$Ca$(e,e'p)$$^{39}$K experiment~\cite{Kramer:1989,Kramer90} (bars). 
   The  peaks in the DOM curves and experimental data correspond to the energies of the quasihole protons in $^{40}$Ca. 
   Note that the experimental fragments in (b) above 4 MeV mostly correspond to $0\textrm{d}5/2$ strength. Figure adapted from Ref.~\cite{Atkinson:2018}.}
   \label{fig:spectral}
\end{figure} 
The $^{40}$Ca DOM self-energy leads to the spectral strength distributions in Fig.~\ref{fig:spectral}. 
   The experimental bars are the results of an angular-momentum decomposition of the experimental spectral function at $T_p$ = 100 MeV as described in Ref.~\cite{Kramer90}.
   The experimental distributions for $\ell = 0,2$ clearly show that the 
   strength is already strongly fragmented at low energies. 
   The main peak in each case represents the valence hole transition of interest.
   This fragmentation is smeared in the DOM via the non-zero imaginary component of the self-energy which is why the DOM curves in Fig.~\ref{fig:spectral} are continuous rather than discrete. The imaginary part of the self-energy approaches zero near $\varepsilon_F$ which results in the sharp peaks of the DOM curve in Fig.~\ref{fig:spectral} (analagous to the what is observed in Fig.~\ref{fig:spectral_n}).   
   The DOM therefore does not yet include the details of the low-energy fragmentation of the valence hole states which requires the introduction of pole structure in the self-energy~\cite{Dickhoff04}.
   The spectroscopic factor of Eq.~(\ref{eq:sf}) corresponds to the main peak of each distribution shown in Fig.~\ref{fig:spectral}. 
   It is calculated directly from the ${}^{40}$Ca DOM self-energy resulting in values of 0.71 and 0.74 for the $0\textrm{d}3/2$ and $1\textrm{s}1/2$ peaks, respectively.
The results are probed in more detail by analyzing the momentum distributions of the $^{40}$Ca$(e,e'p)$$^{39}$K reaction.
 
   In the past, the DWIA calculations by the Nikhef group have been performed using the DWEEPY code~\cite{Giusti:1988}.
   The momentum distributions in Ref.~\cite{Atkinson:2018} are calculated by adapting a recent version of the DWEEPY code~\cite{Giusti:2011} to use the DOM bound-states, distorted waves, and spectroscopic factors as inputs. 
   Before confronting the DOM calculations with the experimental cross sections it is necessary to consider the consequences of the low-energy fragmentation in Fig.~\ref{fig:spectral}.
   For the $0\textrm{d}3/2$ ground state transition (panel (b) of Fig.~\ref{fig:spectral}) there is a clear separation with higher-lying fragments, most of which cannot be distinguished from $0\textrm{d}{5/2}$ contributions as the experiments were not able to provide the necessary polarization information.
In addition, these higher-lying fragments appear to carry little $0\textrm{d}3/2$ strength~\cite{Kramer:2001} , so the DOM spectroscopic factor can therefore be directly used to calculate the cross section of the ground-state peak.

\begin{figure}[t]
      \begin{center}
            \includegraphics[width=0.9\linewidth]{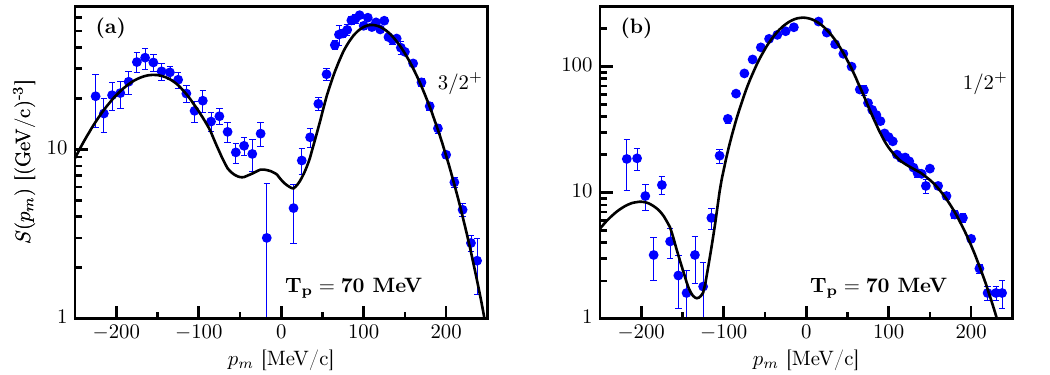}
            \includegraphics[width=0.9\linewidth]{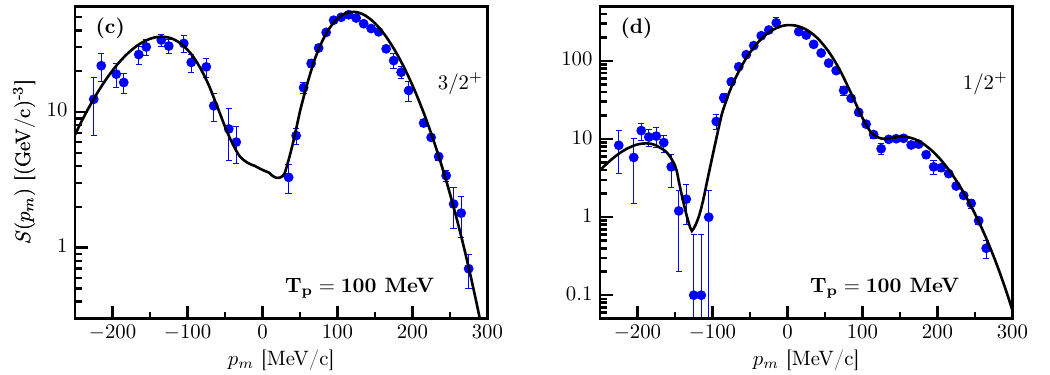}
            \includegraphics[width=0.9\linewidth]{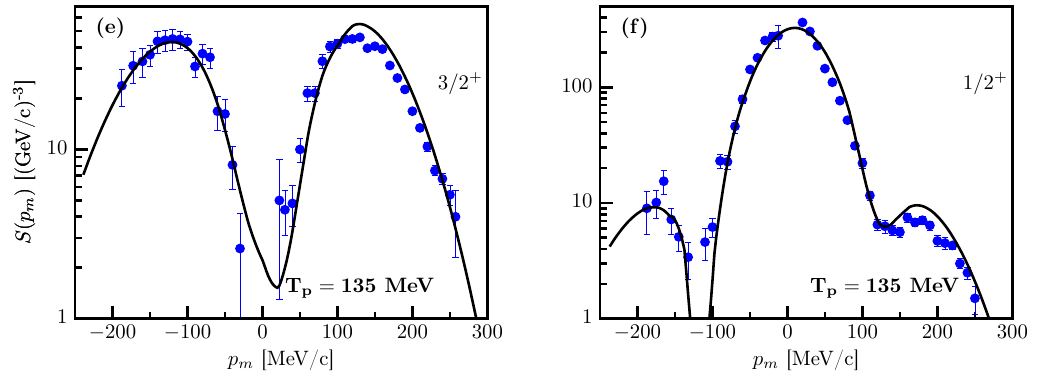}
         \end{center}
      \caption{$^{40}$Ca$(e,e'p)$$^{39}$K spectral functions in parallel kinematics at an outgoing proton kinetic energies of 70,100,135 MeV. The solid line is the calculation using the DOM ingredients, while the points are from the 
      experiment detailed in~\cite{Kramer:1989}.
      (a) Distribution for the removal of the $0\textrm{d}3/2$. The curve contains the DWIA for the $3/2^+$ ground state including a spectroscopic factor of 0.71. (b) Distribution for the removal of the $1\textrm{s}1/2$ proton with a spectroscopic factor of 0.60 for the $1/2^+$ excited state at 2.522 MeV.
       Panels (c) and (e) are the same as (a) except the outgoing proton energy is 100 MeV and 135 MeV, respectively. Panels (d) and (f) are the same as (a) except the outgoing proton energy is 100 MeV and 135 MeV, respectively.
Figure adapted from Ref.~\cite{Atkinson:2018}.
   }
   \label{fig:eep_all}
\end{figure}

The situation is different for the $1\textrm{s}1/2$ distribution which, while dominated by the large fragment at 2.522 MeV, exhibits substantial nearby strength as shown in Fig.~(\ref{fig:spectral})a.
   These contributions come from other discrete poles in the propagator, reflecting the mixing of the $1\textrm{s}1/2$ orbit to more complicated excitations nearby in energy. 
   The origin of these additional discrete poles is not explicitly included in the DOM, although there is a smooth energy-dependent imaginary term in the self-energy to approximate their effect
   on the spectral strength~\cite{Exposed!}. 
   This approximation is sufficient when discussing integrated values such as the charge density and particle number, but falls short when considering details of the low-energy fragmentation into discrete energies as in the present situation. 
   The calculated DOM spectroscopic factor therefore includes strength in the neighborhood of the quasihole energy, resulting in an inflated value. 
   This effect is only noticeable in the $\ell=0$ case because there is a non-negligible amount of strength in the region near the peak.
   We turn to experimental data to account for this effect by enforcing that the ratio between the strength of the peak to the total spectral strength shown in the energy domain of Fig.~(\ref{fig:spectral}) 
   is the same between the data as for the DOM, 
   \begin{equation}
      \frac{\mathcal{Z}_F^{\text{DOM}}}{\int dE\ S^{\text{DOM}}(E)} = \frac{\mathcal{Z}_F^{\text{exp}}}{\int dE\ S^{\text{exp}}(E)}.
      \label{eq:ratio}
   \end{equation}
   Accounting for the contributions to the momentum distribution from different energies by scaling the DOM spectroscopic factor is justified by observing that the shape of the momentum distribution calculated at 
   similar energies is identical, with the strength being the only difference~\cite{Kramer:1989}. 
   The scaling of the spectroscopic factor leads to a reduction from 0.74 to 0.60.
   As mentioned, no correction is needed for the $0\textrm{d}3/2$ spectroscopic factor.
The resulting momentum distributions are shown in Fig.~\ref{fig:eep_all}. The previous analysis of the Nikhef group at $T_p=100$ MeV~\cite{Kramer:1989} produced a comparable reproduction of the data with somewhat smaller spectroscopic factors, as shown in Table~\ref{table}.

 In order to estimate the uncertainty for the DOM spectroscopic factors, we followed the bootstrap method from Ref.~\cite{Varner91} which was also employed in Ref.~\cite{Mahzoon:2017} to assess the uncertainty for the neutron skin in ${}^{48}$Ca.
New modified data sets were created from the original data by randomly renormalizing each angular distribution or excitation function within the experimental error to incorporate fluctuations from the systematic errors. The resulting uncertainties are listed in Table~\ref{table}.

  \begin{table}[b]
   \caption{Comparison of spectroscopic factors deduced from the previous analysis~\cite{Kramer:1989} using the Schwandt optical potential~\cite{Schwandt:1982} to the normalization of the corresponding overlap functions obtained in the present analysis from the DOM including an error estimate as described in the text.}
   \vspace{0.5cm}
   \begin{center}
         \begin{tabular}{ c c c } 
            \hline
            $\mathcal{Z}$ & $0\textrm{d}3/2$ & $1\textrm{s}1/2$\\
            \hline
            \hline
            Ref.~\cite{Kramer:1989} & $0.65 \pm 0.06$ & $0.51 \pm 0.05$\\
            \hline
            DOM & $0.71 \pm 0.04$ & $0.60 \pm 0.03$ \\
            \hline
         \end{tabular}
         \end{center}
   \label{table} 
\end{table}

The DOM results yield at least as good agreement with the data as the standard analysis of Ref.~\cite{Kramer:1989} for the 100 MeV outgoing protons.
The main difference in the description can be pinpointed to the use of nonlocal potentials to describe the distorted waves.
Nonlocal potentials tend to somewhat suppress interior wave functions of scattering states and introduce an additional $\ell$ dependence as compared to local potentials.
We therefore concluded that this consistent treatment clarifies that spectroscopic factors will be larger by about 0.05 when the proper nonlocal dispersive potentials are employed.

The DOM treatment of experimental data associated with both the particle and hole aspects of the single-particle propagator furthermore allows for an assessment of the quality of the DWIA to describe exclusive $(e,e'p)$ cross sections with outgoing proton energies around 100 MeV.
It is therefore fortunate that additional data were obtained at 70 and 135 MeV to further delineate the domain of validity for the DWIA description of the reaction.
We document in Fig.~\ref{fig:eep_all}(a)-(b) the results when DOM ingredients are employed at this lower energy for the two valence hole states in ${}^{39}$K. 
The only difference in the DOM calculations for these cases is the energy of the outgoing proton wave function; the overlap functions and spectroscopic factors remain the same. 

 The agreement with the data at 135 MeV shown in Fig.~\ref{fig:eep_all}(e)-(f) is slightly worse but still acceptable. 
 At this energy (and corresponding value of the electron three-momentum transfer) the contribution of the transverse component of the excitation operator, where other mechanisms contribute in addition to those included in the present operator, will be larger. 
Given these results, it seems that parallel kinematics, in which the longitudinal part of the operator dominates, and a proton energy around 100 MeV, as chosen by the Nikhef group, is optimal for probing the removal probability of valence protons.
We note that this can only be achieved when an analysis is conducted in which all ingredients are provided by a nucleon self-energy that is constrained by all relevant available data as in the DOM.
The excellent agreement found here therefore supports the validity of the DOM approach as it is able to automatically account for the DWIA cross section in the domain where this approximation is expected to be valid.

   \begin{figure}[t]
      \begin{center}
            \includegraphics[scale=0.7]{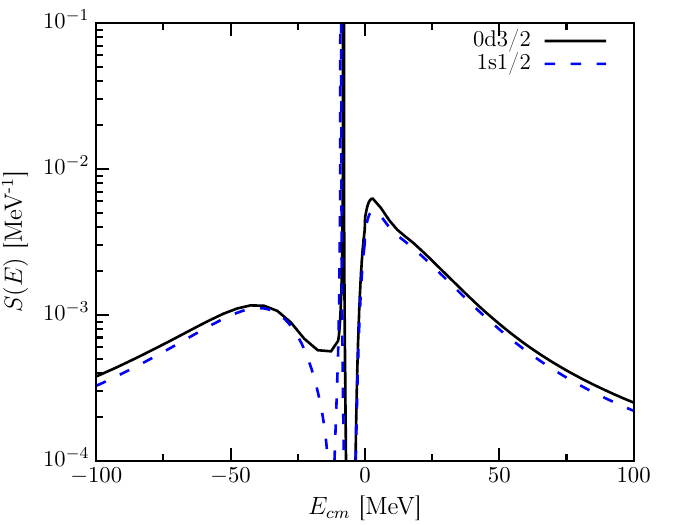}
         \end{center}
         \caption{Spectral distribution of the $0\textrm{d}3/2$ and $1\textrm{s}1/2$ orbits as a function of energy. Additional strength outside this domain is not shown. Figure adapted from Ref.~\cite{Atkinson:2018}.}
   \label{fig:spectralsd}
\end{figure} 
The DOM results also generate the complete spectral distribution for the $0\textrm{d}3/2$ and $1\textrm{s}1/2$ orbits according to 
 \begin{equation}
      S^{n-}_{\ell j}(E) = \sum_{\alpha,\beta}\psi^n_{\ell j}(\alpha)S^h_{\ell j}(\alpha,\beta;E)\psi^n_{\ell j}(\beta) ,
      \label{eq:qh_strength}
   \end{equation}
   and similarly for strength above the Fermi surface~\cite{Dussan:2014}
   \begin{eqnarray}
\!\!\! S_{\ell j}^{n+}(E) 
=  \int \!\! dr r^2 \!\! \int \!\! dr' r'^2 \psi^{n}_{\ell j}(r) S_{\ell j}^{p}(r ,r' ; E) \psi^{n}_{\ell j}(r') ,
\label{eq:specfunc}
\end{eqnarray}
where the actual procedure involves a double integral in coordinate space over the particle spectral amplitude.   
These distributions are displayed in Fig.~\ref{fig:spectralsd} from -100 to 100 MeV.
The energy axis refers to the $A-1$ system below the Fermi energy and the $A+1$ system above.
For clarity, a small imaginary strength in the self-energy near the Fermi energy was employed giving the peaks a small width.
The occupation probabilities are obtained from 
   \begin{equation}
      n^n_{\ell j} = \int_{-\infty}^{\epsilon_F} dES^{n-}_{\ell j}(E),
      \label{eq:occ}
   \end{equation}
   and correspond to 0.80 and 0.82 for the $0\textrm{d}3/2$ and $1\textrm{s}1/2$ orbits, respectively.
The strength at negative energy not residing in the DOM peak therefore corresponds to 9\% and 7\%, respectively.
This information is constrained by the proton particle number and the charge density.
The strength above the Fermi energy is constrained by the elastic-scattering data and generates 0.17 and 0.15 for the $0\textrm{d}3/2$ and $1\textrm{s}1/2$ orbits, respectively, when Eq.~(\ref{eq:depl}) 
\begin{equation}
d^{n}_{\ell j} = \int_{\varepsilon_F}^{\infty} \!\!\!\! dE\ S_{\ell j}^{n+}(E) ,
\label{eq:depl} 
\end{equation}
is employed up to 200 MeV.
The sum rule given by Eq.~(\ref{eq:sumr})
\begin{eqnarray}
\!\! 1 =  n^{n}_{\ell j} + d^{n}_{\ell j} \!\! =\bra{\Psi^A_0} a^\dagger_{n \ell j} a_{n \ell j} +a_{n \ell j}a^\dagger_{n \ell j}  \ket{\Psi^A_0} ,
\label{eq:sumr} 
\end{eqnarray}
associated with the anticommutation relation of the fermion operators, therefore suggests that an additional 3\% of the strength resides above 200 MeV, similar to what was found in Ref.~\cite{Dussan:2014}.
Strength above the energy where surface physics dominates can be ascribed to the effects of short-range and tensor correlations.
The main characterizations of the strength distribution shown in Fig.~55 of Ref.~\cite{Dickhoff04} are therefore confirmed for ${}^{40}$Ca.
The present results thus suggest that it is possible to generate a consistent picture of the strength distributions of these orbits employing all the available experimental constraints.
We therefore conclude that it is indeed quite meaningful to employ concepts like spectroscopic factors and occupation probabilities when discussing correlations in nuclei.

\subsection{$^{48}$Ca$(e,e'p)^{47}$K}
\label{sec:Ca48-eep}
\begin{figure}[h]
   \begin{center}
         \includegraphics[scale=1.0]{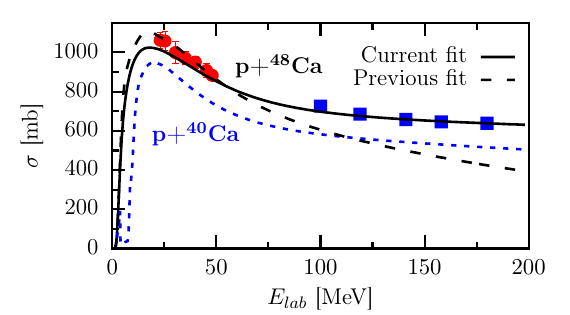}
      \end{center}
   \caption[The proton reaction cross section for $^{48}$Ca.]
   {Proton reaction cross sections for $^{48}$Ca and $^{40}$Ca. The solid line represents the current $^{48}$Ca fit~\cite{Atkinson:2019} while the dashed line depicts the previous $^{48}$Ca fit~\cite{Mahzoon:2017}. The dotted
   line represents the $^{40}$Ca fit from Ref.~\cite{Atkinson:2018}. The circular points are the same $^{48}$Ca experimental data used in Ref.~\cite{Mueller:2011} and were included in the previous fit. The square points
   are extrapolated from the $\sigma_\text{react}^{40}(E)$ experimental data points at the corresponding energies and included in the current $^{48}$Ca fit. Figure adapted from Ref.~\cite{Atkinson:2019}.}
   \label{fig:react48}
\end{figure}

The first DOM fit of $^{48}$Ca was published in Ref.~\cite{Mahzoon:2017}. However, just as in the case of $^{40}$Ca in
Refs.~\cite{Mahzoon:2014,Atkinson:2018}, the proton reaction cross section is underestimated around 200 MeV. While there are no experimental data for $^{48}$Ca at these energies, there is a data point at 700 MeV of the
proton reaction cross section for $^{40}$Ca and $^{48}$Ca~\cite{Anderson700}. Comparing the available data for $\sigma_\text{react}^{40}(E)$ at 200 MeV and 700 MeV reveals that the reaction cross section essentially stays flat
between these energies. It is reasonable to expect that $\sigma_\text{react}^{48}(E)$ assumes the same shape as $\sigma_\text{react}^{40}(E)$
at high energies. Thus, data points are extrapolated from the $^{40}$Ca experimental data at energies above 100 MeV by applying the ratio that is seen in the 700 MeV data for
$\sigma_\text{react}^{48}(E)/\sigma_\text{react}^{40}(E)$~\cite{Atkinson:2019}. The extrapolated points are shown as blue squares in Fig.~\ref{fig:react48} while the updated fit is represented with the solid curve. 
The remainder of the fit did not change significantly from
Ref.~\cite{Mahzoon:2017}.

\begin{figure}[t]
   \begin{center}
         \includegraphics[scale=1.0]{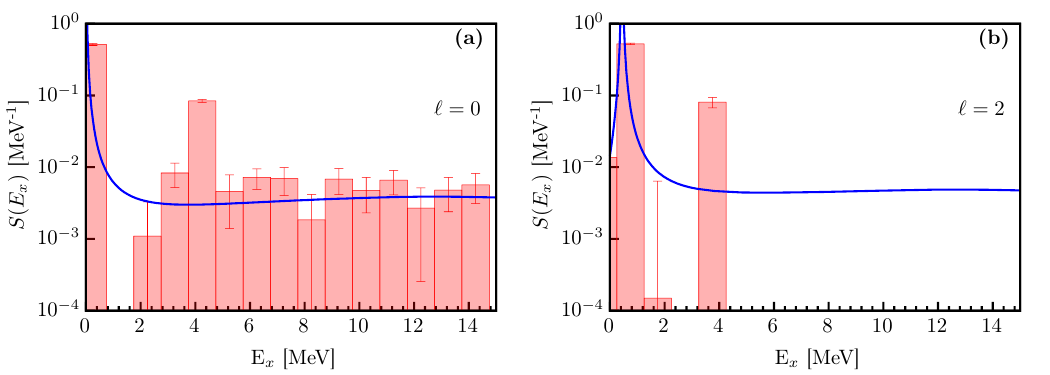}
      \end{center}
   \caption[Spectral strength as a function of excitation energy in $^{48}$Ca]{
      Spectral strength as a function of excitation energy in $^{48}$Ca. The solid lines are DOM spectral functions for (a) the $1\textrm{s}1/2$ and (b) the $0\textrm{d}3/2$ proton orbitals.
      The histograms are the excitation energy spectra in $^{47}$K  extracted from the $^{48}$Ca$(e,e'p)$$^{47}$K experiment~\cite{Kramer:2001,Kramer90}.  The  peaks in the DOM curves and experimental data
      correspond to the quasihole energies of the protons in $^{48}$Ca. The experimental spectrum in (b) is the isolated $0$d$3/2$ orbital. Figure adapted from Ref.~\cite{Atkinson:2019}.
   } 
   \label{fig:spectral_48}
\end{figure}

To analyze the proton spectroscopic factors, the $^{48}$Ca$(e,e'p)^{47}$K cross section is calculated using the DWIA following the same procedure detailed in Sec.~\ref{sec:ca40-eep} for $^{40}$Ca. 
The experimental data of the $^{48}$Ca$(e,e'p)^{47}$K reaction were obtained in parallel kinematics for outgoing proton kinetic energies of $T_p = 100$~MeV at Nikhef
and previously published in Ref.~\cite{Kramer:2001}. Just as in Ref.~\cite{Atkinson:2018}, the DOM spectroscopic factors need to be renormalized by incorporating the observed experimental fragmentation of the strength near the Fermi energy that is not yet included in the DOM self-energy. The experimental strength distributions for the $\ell=0$ and the $\ell=2$ excitations of $^{47}$K are shown in Fig.~\ref{fig:spectral_48}, overlaid with the corresponding DOM spectral
functions calculated from Eq.~\eqref{eq:spec}. Analogously to the $^{40}$Ca calculation, the distributions in Fig.~\ref{fig:spectral_48} are used to renormalize the DOM spectroscopic factors using Eq.~\eqref{eq:ratio}.
This scaling results in a reduction from 0.64 to 0.55 for the $1$s$1/2$ orbital and from 0.60 to 0.58 for the $0$d$3/2$ orbital. These values are in good agreement with
originally published spectroscopic factors~\cite{Kramer:2001}, as seen in Table~\ref{table:sf_48}. The uncertainties in the values of the spectroscopic factors were determined using the same bootstrap method discussed in Sec.~\ref{sec:ca40-eep}. 

\begin{table}[b]
   \begin{center}
         \begin{tabular}{ c c c } 
            \hline
            \hline
            $\mathcal{Z}$ & $0\textrm{d}3/2$ & $1\textrm{s}1/2$\\
            \hline
            Ref.~\cite{Kramer:2001} & $0.57 \pm 0.04$ & $0.54 \pm 0.04$\\
            \hline
            DOM & $0.58 \pm 0.03$ & $0.55 \pm 0.03$ \\
            \hline
            \hline
         \end{tabular}
\end{center}
   \caption[Comparison of spectroscopic factors in $^{48}$Ca]{Comparison of spectroscopic factors in $^{48}$Ca deduced from the previous analysis~\cite{Kramer:2001} using the Schwandt optical potential~\cite{Schwandt:1982} to the normalization of the corresponding overlap functions obtained in the present analysis from the DOM including an error estimate as described in the text.}
   \label{table:sf_48} 
\end{table}

\begin{figure}[t]
   \begin{center}
         \includegraphics[scale=1.0]{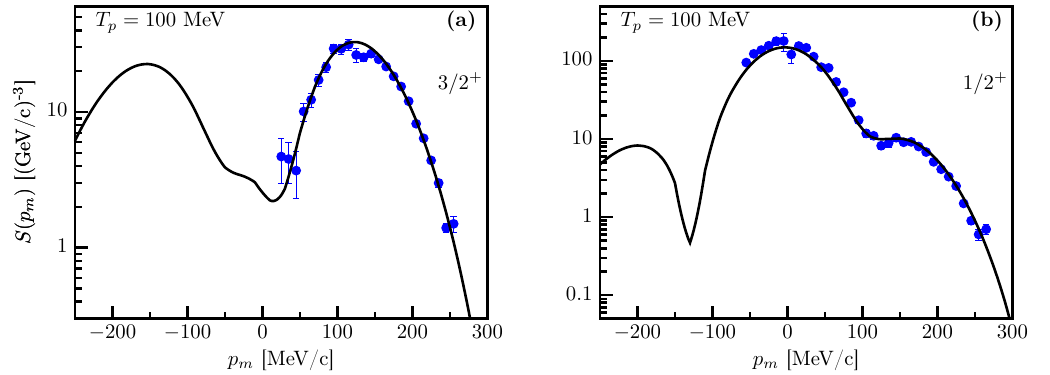}
      \end{center}
   \caption[$^{48}$Ca$(e,e'p)$$^{47}$K spectral functions in parallel kinematics at an outgoing proton kinetic energy of 100 MeV]{
      $^{48}$Ca$(e,e'p)$$^{47}$K spectral functions in parallel kinematics at
      an outgoing proton kinetic energy of 100 MeV. The solid line is the
      calculation using the DOM ingredients while the points are from the
      experiment detailed in Ref.~\cite{Kramer:2001}.  (a) Distribution for the
      removal of the $1\textrm{s}1/2$ proton. The curve contains the
      DWIA for the $1/2^+$ ground state using the DOM generated spectroscopic
      factor of 0.55 (renormalized using Eq.~\eqref{eq:ratio})  (b)
      Distribution for the removal of the $0\textrm{d}3/2$ with a DOM
      generated spectroscopic factor of 0.58 (renormalized using
      Eq.~\eqref{eq:ratio}) for the $3/2^+$ excited state at 0.36 MeV.
      Figure adapted from Ref.~\cite{Atkinson:2019}
}
   \label{fig:eep_48}
\end{figure}

Employing the resulting renormalized spectroscopic factors leads to quantitative agreement with the experimental momentum distributions shown in Fig.~\ref{fig:eep_48}. The comparison of the spectroscopic factors in $^{48}$Ca and $^{40}$Ca, $\mathcal{Z}_{48}$ and $\mathcal{Z}_{40}$, in Table~\ref{table:sf_comp} reveals that both orbitals experience a reduction with the addition of eight neutrons. This indicates that strength from the spectroscopic factors is pulled to the continuum in $S(E)$ when eight neutrons are added to $^{40}$Ca. 
Thus, the stronger coupling to surface excitations in $^{48}$Ca, demonstrated by the larger proton reaction cross section when compared to $^{40}$Ca (see Fig.~\ref{fig:react48}), strongly contributes to the quenching of the proton
spectroscopic factor. It is important to note how crucial the extrapolated high-energy proton reaction cross-section data are in drawing
these conclusions. Without them, there is no constraint for the strength of the spectral function at large positive energies, which could result in no quenching of the spectroscopic factors of $^{48}$Ca due
to the sum rule, Eq.~(\ref{eq:sumr}), that requires the strength to integrate to one when all energies are considered~\cite{Dussan:2014,Exposed!}.

\begin{table}[b]
   \caption{Comparison of DOM spectroscopic factors in $^{48}$Ca and $^{40}$Ca.
   These factors have not been renormalized and represent the aggregate strength near the Fermi energy.}
   \vspace{0.5cm}
   \begin{center}
         \begin{tabular}{ c c c } 
            \hline
            \hline
            $\mathcal{Z}$ & $0\textrm{d}3/2$ & $1\textrm{s}1/2$\\
            \hline
            $^{40}$Ca & $0.71 \pm 0.04$ & $0.74 \pm 0.03$ \\
            \hline
            $^{48}$Ca & $0.60 \pm 0.03$ & $0.64 \pm 0.03$ \\
            \hline
            \hline
         \end{tabular}
         \end{center}
   \label{table:sf_comp} 
\end{table}


In addition to the depletion of the spectroscopic factor due to long-range correlations, strength is also pulled to continuum energies due to SRC.
A large portion of high-momentum content is caused by the tensor force in the nucleon-nucleon ($NN$) interaction. In particular, the tensor force preferentially acts on pairs of neutrons
and protons ($np$ pairs) with total spin $S=1$. This phenomenon is known as $np$ dominance~\cite{Hen:2017}, and is demonstrated by a factor of 20 difference between the number of observed $np$ SRC pairs and
the number of observed $pp$ and $nn$ SRC pairs in exclusive $(e,e'pp)$ and $(e,e'p)$ cross section measurements of $^{12}$C, $^{27}$Al, $^{56}$Fe, and $^{208}$Pb~\cite{Hen:2017}.
The dominance of $np$ SRC pairs would imply that the number of high-momentum protons observed in a nucleus is dependent on how many neutrons it contains. More specifically, one would expect that the
high-momentum content of protons would increase with neutron excess since there are more neutrons available to make $np$ SRC pairs. The CLAS collaboration confirmed this asymmetry dependence by measuring the
high-momentum content of protons and neutrons from $(e,e'p)$ and $(e,e'n)$ cross section measurements in $^{12}$C, $^{27}$Al, $^{56}$Fe, and $^{208}$Pb~\cite{Duer:2018}.

\begin{figure}[t]
   \begin{center}
         \includegraphics[scale=1.0]{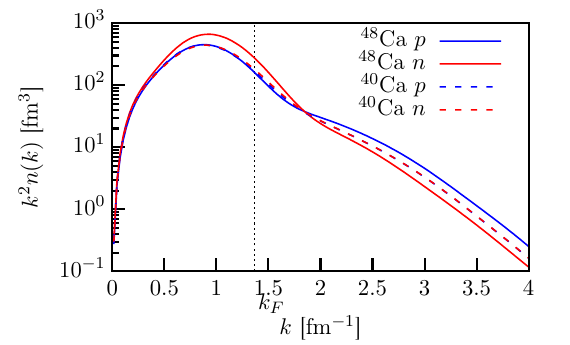}
      \end{center}
   \caption[Comparison of DOM calculated momentum distribution of protons and
   neutrons in $^{48}$Ca and $^{40}$Ca.]
   {
      Comparison of DOM calculated momentum distributions of protons (blue) and
      neutrons (red) in $^{48}$Ca (solid) and $^{40}$Ca (dashed). The dotted line
      marks the value used for $k_F$. Figure adapted from Ref.~\cite{Atkinson:2019}.
   }
   \label{fig:kcomp}
\end{figure} 

This effect can be studied by comparing the DOM generated momentum distributions
for $^{40}$Ca and $^{48}$Ca, since the only difference between them is the eight
additional neutrons in $^{48}$Ca mainly filling the 0f$7/2$ shell. 
It is clear in Fig.~\ref{fig:kcomp} that the $^{48}$Ca proton momentum distribution (solid blue line) has
more high-momentum content than the $^{40}$Ca proton momentum distribution (dashed
blue line). Since the number of protons does not change between
$^{40}$Ca and $^{48}$Ca, the added high-momentum content in the tail of $^{48}$Ca
is accounted for by a reduction of the distribution in the $k<k_F$ region. Turning now to the neutrons in
Fig.~\ref{fig:kcomp} (red lines), the $^{48}$Ca momentum distribution is larger in
magnitude than the $^{40}$Ca distribution for $k<k_F$. This is not
surprising since there are now eight more neutrons which are dominated by low-momentum content. The high-momentum content of the neutrons in $^{40}$Ca decreases from $14.7\%$ to $12.6\%$ when eight neutrons are added to form $^{48}$Ca while the high-momentum content of the protons increases from 14.0\% to 14.6\%.
The effects of the asymmetry of $^{48}$Ca on high-momentum content are evident in the fact that there are more high-momentum protons than neutrons. Both
the increase in proton high-momentum content and the decrease in neutron high-momentum content are qualitatively consistent with the CLAS
measurements of neutron-rich nuclei~\cite{Duer:2018} and support the $np$-dominance picture as predicted in Refs.~\cite{Rios:2009,Rios:2014}. Note that at this stage of the DOM development, no attempt has been made to quantitatively account (i.e. introduce additional constraints) for the CLAS observations.

Another manifestation of the more correlated protons can be seen in the spectral
functions of Fig.~\ref{fig:spectral_n}. The broader peaks
of the proton spectral functions in Fig.~\ref{fig:spectral_n}(a) compared to those of the neutrons in Fig.~\ref{fig:spectral_n}(b) indicate that the protons are more correlated.
Furthermore, increased proton high-momentum content in $^{48}$Ca is a result from the added strength in the continuum of the hole spectral function when compared to that of $^{40}$Ca. 
To conserve proton number (and preserve the sum rule of Eq.~(\ref{eq:sumr})), an
increase in strength at continuum energies in $S_{\ell j}(E)$ of $^{48}$Ca must be
compensated by a decrease in strength from energies close to the proton Fermi energy
in $^{48}$Ca.  In particular, this contributes to the quenching of the spectroscopic
factors of the $0$d$3/2$ and $1$s$1/2$ orbitals, before
renormalization (see Eq.~\eqref{eq:ratio}), in $^{48}$Ca from the values for
$^{40}$Ca as can be seen in Table~\ref{table:sf_comp}.  In this way, the
spectroscopic factor provides a link between the low-momentum knockout experiments
done at Nikhef and the high-momentum knockout experiments done at JLAB by the CLAS
collaboration.

The success of the DOM in describing both $^{40}$Ca$(e,e'p)^{39}$K and $^{48}$Ca$(e,e'p)^{47}$K results has provided a foothold for understanding the quenching of spectroscopic factors. A DOM investigation across the nuclear chart would deepen our understanding, as a data-informed spectroscopic factors could be generated for each nucleus using Eq.~\eqref{eq:sf}. This would require a global parametrization of the DOM which is currently in development. In the meantime, 
we note that for $^{208}$Pb (see Ref.~\cite{Atkinson:2020} for fit), the DOM values of the valence spectroscopic factors are consistent with the observations of Ref.~\cite{Lichtenstadt79} and the interpretation of Ref.~\cite{Vijay84}.
The past extraction of spectroscopic factors using the $^{208}$Pb$(e,e'p)^{207}$Tl reaction yielded a value around 0.65 for the valence 
$\mathrm{2s_{1/2}}$ orbit~\cite{Ingo91} based on the results of Ref.~\cite{Quint86,Quint87}.
While the use of nonlocal optical potentials may slightly increase this value as shown in Ref.~\cite{Atkinson:2018}, it may be concluded that the value of 0.69 obtained from the DOM analysis is consistent with the past result.
Nikhef data obtained in a large missing energy and momentum domain~\cite{Batenburg01} can now be consistently analyzed employing the complete DOM spectral functions.

\section{Proton-induced Knockout}
\label{sec:p2p}

As discussed in Sec.~\ref{sec:intro}, knockout reactions can be induced by nuclear projectiles such as protons. While these reactions are not as clean due to the probe interacting through the nuclear $pp$ force rather than the electromagnetic $ep$ force, the DWIA description does a fairly good job of reproducing experimental data. Furthermore, these reactions are not limited to forward kinematics like their electron-induced counterpart - the proton can be either the beam or the target. This is useful because it allows for the study of nuclei far from stability by utilizing rare isotope beams in labs such as the DOE flagship facility for rare isotope beams (FRIB). 

Since we have an accurate description of $^{40}$Ca$(e,e'p)^{39}$K using the DOM ingredients, we are in a good position to investigate the reaction description of the analogous $^{40}$Ca$(p,2p)^{39}$K reaction. The kinematics of the $(p,2p)$ experiment are setup in a similar manner to those of the $(e,e'p)$ experiment and the outgoing proton energy is 100 MeV, which we showed is an optimal energy for a good DWIA description of knockout~\cite{Atkinson:2018}. While the experiment we compare to was performed using a proton beam on a stable $^{40}$Ca target~\cite{Noro:2023}, it serves as a benchmark for the DWIA description of proton-induced knockout so it can be applied in more exotic cases when protons are used as targets for rare isotope beams. 

\subsection{DWIA in $(p,2p)$}

In Ref.~\cite{Yoshida:2021}, the factorized form of the nonrelativistic DWIA with the spin degrees of freedom is employed.
The transition matrix $T$ within the distorted wave impulse approximation framework
is given by
\begin{align}
T_{\mu_1 \mu_2 \mu_0 \mu_j}
=
\sum_{\mu'_1 \mu'_2 \mu'_0 \mu_p}
\tilde{t}_{\mu'_1 \mu'_2 \mu'_0 \mu_p} 
\int d\bm{R}\,
\chi_{1, \mu'_1 \mu_1}^{(-)*}(\bm{R})
\chi_{2, \mu'_2 \mu_2}^{(-)*}(\bm{R})&
\chi_{0, \mu'_0 \mu_0}^{(+)}(\bm{R})
e^{-i\alpha_{R}\bm{K}_0 \cdot\bm{R}} \nonumber\\ 
\times
&\sum_{m}
\left(\ell ms_p\mu_p | j\mu_j\right)
\psi^n_{\ell jm}(\bm{R}).
\label{eq:tmatrix}
\end{align}
The incident and two emitted protons are labeled as particle $0$--$2$,
while the bound proton in the initial state is labeled as $p$.
$\chi_{i,\mu'_i \mu_i}$ is a distorted wave of particle $i=0,1,2$
having the asymptotic (local) third component $\mu_i$ ($\mu'_i$) of its spin $s_i = 1/2$.
The outgoing and incoming boundary conditions of the distorted waves 
are denoted by superscripts $(+)$ and $(-)$, respectively.
$\bm{K}_0$ is the momentum (wave number) of the incident proton and 
$\alpha_{R}$ is the mass ratio of the struck particle and the target.
$n$ is the radial quantum number, and $\ell, j, m$ are the single-particle
orbital angular momentum, total angular momentum, and third component of $\ell$,
respectively.
$\psi^n_{\ell jm}$ is the single-particle wave function (SPWF) normalized to unity.
$\tilde{t}_{\mu'_1 \mu'_2 \mu'_0 \mu_p}$ is the 
matrix element of the $pp$ effective interaction $t_{pp}$:
\begin{align}
\tilde{t}_{\mu'_1 \mu'_2 \mu'_0 \mu_p} 
&=
\Braket{
\bm{\kappa}', \mu'_1 \mu'_2 
| t_{pp} |
\bm{\kappa}, \mu'_0 \mu_p
},
\label{eq:tpp}
\end{align}
where $\bm{\kappa}$ and $\bm{\kappa}'$ are relative momenta of two protons
in the initial and the final states, respectively.
The factorization procedure of $t_{pp}$ is explained using the local semi-classical approximation
(LSCA) and the asymptotic momentum approximation (AMA) in the appendix of Ref.~\cite{Yoshida:2021}.
It should be noted that the factorized DWIA is often regarded as a result of the zero-range approximation
but $t_{pp}$ is a finite-range interaction 

The triple differential cross section (TDX) with respect to the 
emitted proton energy $T_1^\mathrm{\mathrm{lab}}$ and emission angles 
$\Omega_1^\mathrm{\mathrm{lab}}$ and $\Omega_2^\mathrm{\mathrm{lab}}$
is given as
\begin{align}
\frac{d^3\sigma^{\mathrm{\mathrm{lab}}}}{dT_1^\mathrm{\mathrm{lab}} d\Omega_1^\mathrm{\mathrm{lab}} d\Omega_2^\mathrm{\mathrm{lab}}}
=
\mathcal{Z}^n_{lj}
\mathcal{J}_\mathrm{\mathrm{lab}G} F_\mathrm{kin}
\frac{(2\pi)^4}{\hbar v_\alpha}
\frac{1}{(2s_0+1)(2j+1)}
\sum_{\mu_1 \mu_2 \mu_0 \mu_j} 
\left| T_{\mu_1 \mu_2 \mu_0 \mu_j} \right|^2,
\label{eq:tdx}
\end{align}
with $ \mathcal{Z}^n_{lj}$, $\mathcal{J}_\mathrm{\mathrm{lab}G}$, $F_\mathrm{kin}$, $v_\alpha$ being 
the spectroscopic factor,
the Jacobian 
from the center-of-mass frame to the Laboratory frame, kinetic factor, and
the relative velocity of the incident proton and the target, respectively.
Quantities with superscript $\mathrm{lab}$ are evaluated in the laboratory frame 
while the others are in the center-of-mass frame.
See Sec. 3.1 of Ref.~\cite{Wakasa17} for details.

Equations ~\eqref{eq:current}, ~\eqref{eq:tpp}, and ~\eqref{eq:tdx} for electron- and proton-induced knockout, respectively, have many similarities. They both employ the same spectroscopic factor, bound-state wave function, and 100 MeV outgoing proton distorted wave. The proton-induced expression includes two addition proton distorted waves to account for the incoming and outgoing projectile proton, but the main difference between these two equations is the appearance of the $pp$ interaction in the form of $\tilde{t}_{\mu'_1 \mu'_2 \mu'_0 \mu_p}$ in Eq.~\eqref{eq:tpp}. In the electron case, this is factorized outside of the hadronic part of the cross section which is not possible in the proton-induced case. We probe this difference by employing the same DOM ingredients between $(e,e'p)$ and $(p,2p)$ cases. 

\subsection{Results and discussion}
\label{sec_result}

\begin{figure}[t]
   \begin{center}
\includegraphics[width=0.49\textwidth]{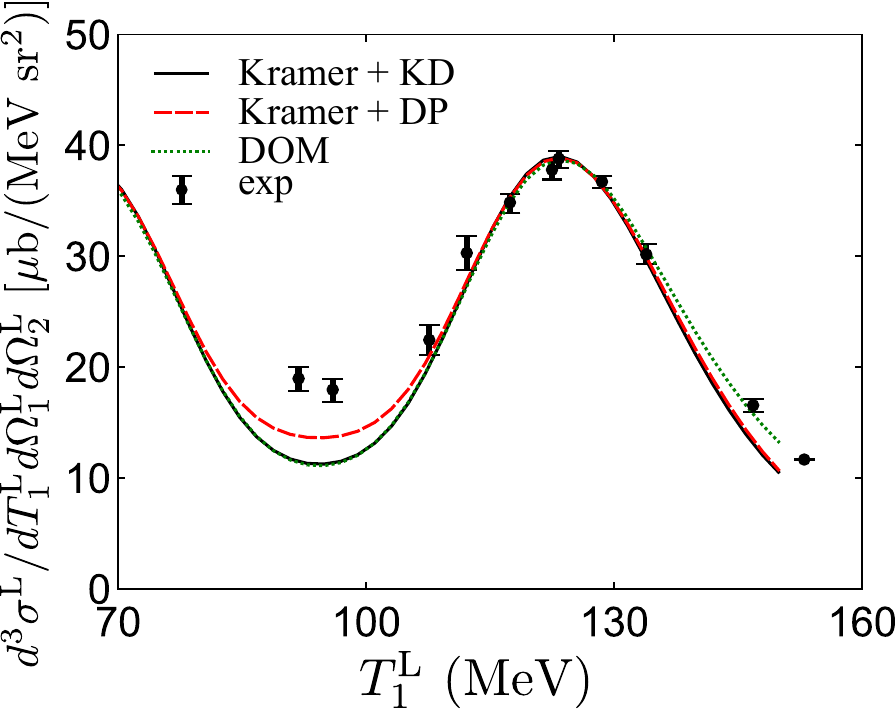}
\includegraphics[width=0.49\textwidth]{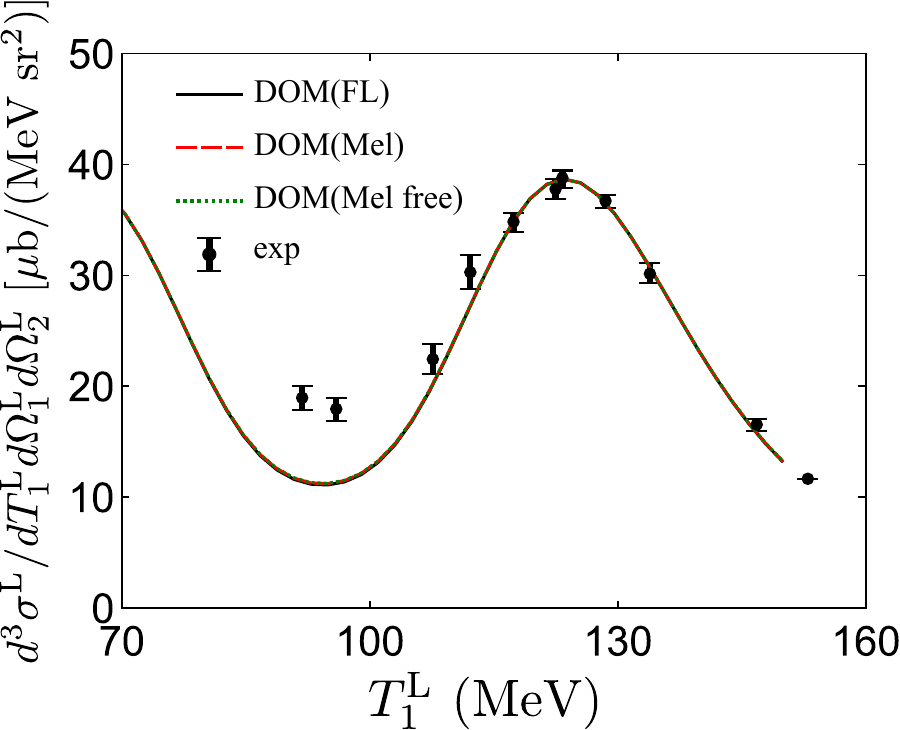}
\end{center}
\caption{ (a) TDX with different optical potentials.
The solid and dashed lines are TDXs with the Koning-Delaroche
optical potential (KD) and Dirac phenomenology (DP),
respectively.
The result with DOM ingredients is also shown as the dotted line. All results reflect cross sections that are normalized with the spectroscopic factors shown in Table~\ref{tab:s-factor}.
(b) TDX calculated using the DOM with different $pp$ effective interactions.
The solid, dashed and dotted lines are TDXs with the 
Franey-Love effective interaction~\cite{Franey85} (FL), 
Melbourne $g$-matrix interaction at mean density~\cite{Amos00} (Mel),
and that at zero density (Mel free),
respectively. See Table~\ref{tab:s-factor} for corresponding normalization (spectroscopic) factors. 
The experimental data taken by the E258 experiment at RCNP~\cite{Noro:2023} 
are also shown.
Figure adapted from Ref.~\cite{Yoshida:2021}.
}
\label{fig:tdx_dw}
\end{figure}

Theoretical knockout cross sections are calculated using the 
DWIA framework with the DOM SPWF and distorted waves.
The reaction kinematics is in a coplanar kinematics and 
the opening angles of the emitted protons are
fixed at the same angle; 
$\phi_1^\mathrm{L} = 0^\circ$, 
$\phi_2^\mathrm{L} = 180^\circ$,
and
$\theta_1^\mathrm{L} = \theta_2^\mathrm{L} = 42.0^\circ$
in the Madison convention~\cite{madison_convention}.
The kinematics of the three particles is then uniquely determined
by $T_1^\mathrm{L}$.
The DOM-DWIA result is compared with those of the phenomenological 
SPWF and the optical potential in panel (a) of Fig~\ref{fig:tdx_dw}.
For this comparison, the DOM-DWIA cross section is adjusted to the data rather than employing the DOM-generated spectroscopic factor from Eq.~\eqref{eq:sf}.
The phenomenological SPWF 
suggested by Kramer \textit{et al.}~\cite{Kramer:2001} and the
Koning-Delaroche optical potential parameter set (KD)~\cite{Koning:2003} 
as well as the
Dirac phenomenology (DP)~\cite{Hama90,Cooper93,Cooper09} are also considered.
Spectroscopic factors are therefore extracted from the ratio of the present calculations and the 
experimental data taken by the E258 experiment at the RCNP~\cite{Noro:2023}
by minimizing
\begin{align}
  \chi^2(\mathcal{Z}_{0d_{3/2}})
  &=
  \sum_i
  \frac{
  \left(\mathcal{Z}_{0d_{3/2}} \sigma_i^{\textrm{DWIA}}-\sigma_i\right)^2
  }{
  \delta_i^2},
  \label{eq:chi}
\end{align}
where $\sigma_i^{\textrm{DWIA}}$ and $\sigma_i$ are theoretical and experimental 
cross sections at data points $i$, respectively, and $\delta_i$ is the associated error of the experimental data.
Obtained spectroscopic factors are summarized in Table~\ref{tab:s-factor}. 
Following Ref.~\cite{Wakasa17}, only the data points around the peak, larger than $25~\mu$b/(MeV sr$^2$),
are fit to reduce the uncertainty.
\begin{table}[bp]
\caption{ Normalization (spectroscopic) factors extracted in $^{40}$Ca($p,2p)^{39}$K using Eq.~\eqref{eq:chi}.}
\begin{center}
\begin{tabular}{lllc}
\hline
\hline
SPWF   & Optical pot. & $pp$ int. & $\mathcal{Z}_{0d_{3/2}}$ \\
\hline
Kramer & KD           & FL           & $0.623   \pm 0.006$    \\
\hline
Kramer & Dirac        & FL           & $0.672   \pm 0.006$    \\
\hline
DOM    & DOM          & FL           & $0.560   \pm 0.005$    \\
\hline
DOM    & DOM          & Mel          & $0.489   \pm 0.005$    \\
\hline
DOM    & DOM          & Mel (free)   & $0.515   \pm 0.005$    \\
\hline
\hline
\end{tabular}
\end{center}
\label{tab:p2p_sf}
\end{table}

The spectroscopic factors obtained from the phenomenological ($p$,$2p$) analysis (the first two rows of Table~\ref{tab:p2p_sf})
are consistent with the phenomenological ($e$,$e'p$) analysis which resulted in 
$0.65 \pm 0.06$~\cite{Kramer:1989}.
On the other hand, the spectroscopic factor obtained using the DOM wave functions to reproduce the ($p$,$2p$) cross section is in disagreement with the DOM  value (using Eq.~\eqref{eq:sf}) of $0.71 \pm 0.04$.
Since the spectroscopic factor is a property of the quasihole bound state, it should not depend on the reaction mechanism or beam energy~\cite{Radici:2002}. 
As shown in Ref.~\cite{Wakasa17}, the spectroscopic factors for the valence levels near the Fermi energies of stable nuclei extracted from ($p,2p$) reactions above 200~MeV, using the DWIA with local potentials, are consistent with those from ($e,e'p$) with uncertainties ranging from 10\% to 15\%. The nonlocality correction to the SPWF and distorted waves is considered to be a primary source of uncertainty in determining these spectroscopic factors~\cite{Wakasa17}. 

Employing different potentials to generate the proton scattering and bound-state wave functions muddies the interpretation of these results. The DOM ingredients are both fully consistent within the DWIA framework and equivalent between the $(e,e'p)$ and $(p,2p)$ reactions. However, $\mathcal{Z}_{0\mathrm{d}3/2}$ obtained with the DOM-DWIA analysis of the $^{40}$Ca($p,2p$) data at 200~MeV, in which the nonlocality is treated in a sophisticated manner,  differs  by
at least 21\%
from the value used to reproduce $(e,e'p)$ data utilizing the same SPWF and proton distorted wave from the DOM. With the ingredients of the DWIA description eliminated as causes for this discrepancy (because of the consistency and commonality between the $(e,e'p)$ and $(p,2p)$ cases), we explore differences in the reaction descriptions to uncover the source of discrepancy.

We first consider the consequences of using three distorted proton waves in the ($p$,$2p$) reaction as compared to just one in the ($e$,$e'p$).
There is an uncertainty associated with the DOM distorted waves due to the experimental data points used in the DOM fit. Recalling the strong correlation between the proton reaction cross sections and the $(e,e'p)$ cross sections demonstrated in Sec.~\ref{sec:DWIA}, we look to uncertainties in the experimental proton reaction cross section data points in energy regions corresponding to those of the distorted proton waves to get a rough estimate of the uncertainty associated with the DOM distorted waves. The proton reaction cross section data points from Refs.~\cite{ca40react-1,ca40react-2} suggest an uncertainty in the corresponding DOM distorted waves around 3\%. Furthermore, due to the kinematics of the reaction, one of the proton energies is as low as 36 MeV. In the DOM analysis of $^{40}$Ca$(e,e'p)^{39}$K, the description of the experimental cross section for outgoing proton energies of 70 MeV, the lowest of the considered proton energies, is somewhat less satisfactory~\cite{Atkinson:2018}. This indicates that the impulse approximation may not be applicable at proton energies of 70 MeV and below. Since one of the outgoing proton energies in this $^{40}$Ca$(p,2p)^{39}$K reaction is even less than 70 MeV, it is reasonable to expect some discrepancy in the $^{40}$Ca$(p,2p)^{39}$K TDX. 
This discrepancy may be reduced when higher proton beam energies are considered but this implies that the DOM analysis has to be extended to higher energies. Noting that previous analyses of $(p,2p)$ and $(e,e'p)$ resulted in consistent spectroscopic factors, we conclude that any inaccuraccies caused by low-energy protons do not explain the large 21\% discrepancy we are observing between DOM descriptions of $(e,e'p)$ and $(p,2p)$. 
We also investigated the uncertainty arising from a 
different choice of the $pp$ effective interactions when employing the DOM in the DWIA.
Three different types of the $pp$ effective interaction,
the Franey-Love effective interaction (FL)~\cite{Franey85},
the Melbourne $g$-matrix interaction at mean density (Mel)~\cite{Amos00},
and that at zero density  (Mel free) were utilized.
The Franey-Love interaction is a free-space $t$-matrix aimed at reproducing high-energy $pp$ scattering cross sections. The Melbourne interactions utilize the so-called $g$-matrix which is an approximation to account for the fact that the $pp$ interaction in $(p,2p)$ takes place in a nucleus rather than a vacuum. The $g$-matrix is typically calculated from the $pp$ interaction via ladder diagrams in infinite nuclear matter and mapped to finite nuclei using the density~\cite{Amos00}.
The mean density of the $(p,2p)$ reaction is defined in Sec. 6.1. of Ref.~\cite{Wakasa17}.
The choice of the $pp$ effective interaction leaves the shape of the TDX unaltered (see panel (b) of Fig.~\ref{fig:tdx_dw}) but changes the magnitude of the TDX such that the normalization factor to reproduce experimental data varies (see Table~\ref{tab:p2p_sf}). 

The uncertainty due to the choice of the $pp$ effective interaction 
results in $\mathcal{Z}_{0d_{3/2}} = 0.489$--$0.560$ which is still inconsistent with the DOM $(e,e'p)$ results~\cite{Atkinson:2018}. 
However, the variation in the spectroscopic factors using the different interactions (see Table.~\ref{tab:s-factor}) indicates that the $(p,2p)$ reaction is sensitive to the chosen effective $pp$ interaction. 
We note that the main difference between $(e,e'p)$ and $(p,2p)$ is the need to employ an in-medium $pp$ interaction which is not well-constrained.
We therefore hypothesize that the $(p,2p)$ reaction must be investigated with a more sophisticated treatment of the $pp$ interaction beyond the standard $t$- or $g$-matrix approach.
One immediate concern is that present treatments of this effective interaction do not allow for energy transfer in the elementary process.
Since a substantial excitation energy is involved in the ($p,2p$) reaction, it implies that the mediators of the strong interaction, in particular the pion, must be allowed to propagate~\cite{Jonathan11}. 
The in-medium effective $pp$ interaction should be calculated in finite nuclei, which can be achieved by utilizing DOM propagators. The formalism for this nucleus-dressed interaction is analogous to that of the $g$-matrix, but instead of mapping the infinite nuclear matter propagator to $^{40}$Ca via the density we can explicitly employ the DOM propagator (Eq.~\eqref{eq:dyson}) of $^{40}$Ca. The incorporation of the $^{40}$Ca DOM propagator in the effective $pp$ interaction should contribute to improving the reaction description such that the DOM spectroscopic factor of $\mathcal{Z}_{0d_{3/2}} = 0.71$ will ultimately describe both $^{40}$Ca$(e,e'p)^{39}$K and $^{40}$Ca$(p,2p)^{39}$K cross sections.

\section{Conclusions}
\label{sec:conclusions}

We have reviewed a nonlocal dispersive optical-model analysis of ${}^{40}$Ca and ${}^{48}$Ca in which we fit elastic-scattering angular distributions, absorption and total cross sections, single-particle energies, charge densities, ground-state binding energies, and particle numbers. When sufficient data are available to constrain the self-energy, the DOM can provide accurate predictions. In particular, the unique capability of the DOM to simultaneously describe bound-state and scattering wave functions leads to fully-consistent DWIA descriptions of knockout reactions. After updating the high-energy reaction cross sections used to constrain the DOM self-energies in $^{40}$Ca and $^{48}$Ca, the predictions $^{40}$Ca$(e,e'p)^{39}$K and $^{48}$Ca$(e,e'p)^{47}$K reproduced the Nikhef experimental data which resulted in the updated spectroscopic factors for both $^{40}$Ca and $^{48}$Ca (see Table~\ref{table:sf_comp})~\cite{Atkinson:2018,Atkinson:2019}. Furthermore, we observe a reduction in the spectroscopic factors from $^{40}$Ca to $^{48}$Ca which is consistent with the quenching observed in the systematic analysis of Ref.~\cite{Aumann21}. Through the spectral-function picture of the nucleus provided by the DOM, we connect the quenching of spectroscopic factors to the increase in high-momentum content of protons when eight neutrons are added to $^{40}$Ca to form $^{48}$Ca. 

The DOM-DWIA description of the proton-induced knockout from $^{40}$Ca, however, does not currently fit in the consistent story of its electron-induced counterpart. Indeed, the DOM-DWIA overestimates the $^{40}$Ca$(p,2p)^{39}$K by 21\% even though the same DOM ingredients are employed which were so successful in describing the $^{40}$Ca$(e,e'p)^{39}$K reaction.
We hypothesize that the largest cause of this discrepancy is the fact that the probe in $(p,2p)$ interacts with the nucleus through the nuclear $pp$ interaction rather than the electromagnetic $ep$ interaction in $(e,e'p)$. 
We are therefore working on improving the $(p,2p)$ description by utilizing DOM propagators to explicitly treat the $pp$ interaction as scattering inside $^{40}$Ca rather than in free-space ($t$-matrix) or infinite nuclear matter ($g$-matrix).  
The ability of the DOM to provide both bound and scattering nucleon wave functions is opening the door to a new research opportunity for the nucleon-nucleon scattering process in many-body systems.
This is of particular importance as nucleus-induced reactions, which utilize the $NN$ interaction in their theoretical description (including $(p,2p)$), can be employed in inverse kinematics to study nuclei off stability at RIB facilities~\cite{Altar:2018,Kawase:2018}. 
There is therefore a clear need to pursue an improved description of the effective interaction in the medium which will also depend on the nucleon asymmetry that is studied in exotic systems.

\section*{Conflict of Interest Statement}

The authors declare that the research was conducted in the absence of any commercial or financial relationships that could be construed as a potential conflict of interest.

\section*{Author Contributions}

The authors provided approximately equivalent contributions to the conceptual and practical aspects of this review.

\section*{Funding}
This work was performed under the auspices of the U.S. Department of Energy by Lawrence Livermore National Laboratory under Contract DE-AC52-07NA27344 and was supported by the LLNL-LDRD Program under Project No. 24-LW-062. This work was also supported by the U.S. National Science Foundation under grants PHY-1912643 and PHY-2207756.

\section*{Acknowledgments}
The authors acknowledge the important contributions to some of this research from Henk Blok, Bob Charity, Louk Lapik\'{a}s, Kazuyuki Ogata, and Kazuki Yoshida.

\bibliographystyle{Frontiers-Vancouver} 
\bibliography{dom-review}

\end{document}